# Twin Supercoil Domain couples the dynamics of molecular motors and plectonemes during bacterial DNA transcription and replication

Marc JOYEUX [*]

*Laboratoire Interdisciplinaire de Physique,*

*CNRS and Université Grenoble Alpes,*

*38400 St Martin d'Hères,*

*France*

Corresponding Author

*(M.J.) E-mail: marc.joyeux@univ-grenoble-alpes.fr.

ORCID : Marc Joyeux: 0000-0002-6282-1846

**ABSTRACT** : The genomic DNA of most bacteria is significantly underwound, which constrains the DNA molecule to adopt a branched plectoneme geometry. Moreover, biological functions like replication and transcription require that the two DNA strands be transiently opened, which generates waves of positive (respectively, negative) supercoiling downstream (respectively, upstream) of the molecular motor, a feature known as «Twin Supercoiled Domain» (TSD). In this work, we used coarse-grained modeling and Brownian Dynamics simulations to investigate the interactions between a TSD and the plectonemes of bacterial DNA. Simulations indicate that the slithering dynamics of short plasmids is not significantly affected by a TSD. In contrast, the TSD potently stimulates the spontaneous displacement modes (diffusion and growth/shrinkage) of the plectonemes of longer DNA molecules. This results in the motor trailing a growing plectoneme behind itself if it translocates more slowly than the maximum slithering speed of plectonemes. In contrast, if the motor translocates more rapidly than this limit, then quasi immobile plectonemes nucleate almost periodically upstream of the motor, grow up to several kbp, detach from the motor, shrink and disappear. The effect of an eventual static bend imposed by the motor and of topological barriers was also investigated.

**INTRODUCTION**

The fact that the genomic DNA molecule of most bacteria is circular (closed) has the important geometrical consequence that the number of times the two strands wind around each other, called the linking number ($Lk$), remains constant as long as the integrity of the two strands is preserved [1-4]. In torsionally relaxed B-DNA, the two strands wind around each other every 10.5 base pairs, meaning that the linking number of a relaxed circular B-DNA molecule with $N$ base pairs (bp) is $Lk_0 = N/10.5$. A circular DNA molecule with $Lk \neq Lk_0$ is said to be supercoiled and the ratio $\sigma = (Lk - Lk_0)/Lk_0$ is known with different names, including superhelical density, density of supercoiling, and specific linking difference (the former one is used in the present paper). Underwound DNA ($Lk < Lk_0$) has negative superhelical density ($\sigma < 0$), whereas overwound DNA ($Lk > Lk_0$) has positive superhelical density ($\sigma > 0$). The linking number $Lk$ can be partitioned into twist ($Tw$) and writhe ($Wr$), such that $Lk = Tw + Wr$ [5]. $Tw$ is the signed number of times the two strands twist around their center line and $Wr$ the signed number of times the center line crosses itself [6]. It is generally considered that the mean ratio of writhe to linking number difference, $\langle Wr \rangle / (Lk - Lk_0)$, is close to 70% in B-DNA [7]. However, salt concentration may let this ratio vary from 50% to 90% [8] and external constraints may also result in twist converting to writhe, or conversely, because neither $Tw$ nor $Wr$ need remain constant. In addition, several fundamental biological functions result in strong perturbations of the local superhelical density. For example, the transcription of genes into messenger RNA (mRNA) molecules, which is catalyzed by RNA polymerase enzymes (RNAP) [9,10] translocating along the DNA molecule [11-15], requires the local opening of the two strands [16]. During transcription elongation, the RNAP consequently sends waves of positive supercoiling downstream of its position and negative supercoiling upstream of it [17-23], generating a so-called "Twin Supercoiled Domain" (TSD) [17-19]. Although much less documented, the replication of circular DNA results similarly in the generation of TSDs, because the two replisomes, which travel around the DNA molecule in opposite directions, also open locally and sequentially the two strands [24-27]. In the rest of this paper, the word "motor" will indiscriminately refer to molecular assemblies which generate a TSD during their translocation along DNA molecules.

Several modeling/simulation works investigated the role of TSDs on the regulation of the transcription of neighboring genes [28-33], the formation of self-interacting domains [34,35], and loop extrusion [35,36]. Other studies dealt with the dynamics of TSDs themselves [37-39]. In particular, we recently investigated the dynamics of TSDs in 21600 bp torsionally relaxed ($\sigma = 0$) and negatively supercoiled ($\sigma = -0.06$) DNA molecules [39] using a coarse-grained model where the action of the motor is modeled as an external torque, in the spirit of [33,34,40,41]. Brownian Dynamics (BD) simulations performed with this model confirmed that the rapid injection of twist in torsionally relaxed DNA results in positively supercoiled plectonemes forming as far as 5000 bp downstream of the motor [38,39]. Moreover, simulations highlighted the fact that positively supercoiled plectonemes never form ahead of motors which translocate along DNA molecules with physiological negative supercoiling ($\sigma = -0.06$), whereas negatively supercoiled ones do form at their upstream side [39].

The present work elaborates on the preliminary results in [39] and aims at providing a comprehensive picture of how the TSD generated by a motor which translocates along a bacterial DNA molecule influences the dynamics of its plectonemes. Indeed, the DNA of most bacteria is negatively supercoiled ($\sigma \approx -0.06$ [42]) and forms branched plectonemes [43,44], which alleviate part of the torsional stress arising from unwinding. The goal is to understand how the waves of positive and negative supercoiling generated by transcription and replication interact with these plectonemes and the extent to which they impact their dynamics. Moreover, we asked ourselves how a certain number of factors may affect these interactions, namely

- Do the length of the DNA molecule and the number of branching points [43,44] affect these interactions and, in particular, does a TSD have the same impact on short plasmids and longer (genomic) DNA molecules ?

- an RNAP enzyme transcribes on average about 50 nucleotides per second [45-47], whereas a replisome copies on average about 500 bases per second [26,27]. Does the translocation speed of the motor have some impact on its interactions with plectonemes ?

- it is known that static bends and base pair mismatches have the ability to pin plectonemes [21,48,49]. Does the bend of about 40° imposed by RNAPs to the DNA molecule [16,50] affect significantly the dynamics of the plectonemes ?

- diffusion of twist along DNA molecules is often blocked by topological barriers, like pairs of LacI repressor proteins [51]. What fingerprint of plectoneme dynamics is left in supercoiled DNA when the diffusion of twist is blocked and motors have stalled ?

This work complements bulk experiments [17-19,45,52] and single-molecule ones [21,23,28] for understanding the relation between DNA supercoiling and biological function during DNA transcription and replication [53].

**THEORETICAL METHODS**

The coarse-grained model used in the present work has been described in detail in [39]. Its main features are sketched here briefly for the sake of clarity.

DNA molecules are modeled as circular chains of $n=600$ or $n=2880$ beads with radius $a=1.0$ nm separated at equilibrium by a distance $l_0 = 2.5$ nm. Each bead represents 7.5 bp, so that chains composed of 600 and 2880 beads represent DNA molecules with 4500 bp and 21600 bp, respectively. Associated to each bead $k$ are a vector $\mathbf{r}_k$, which describes its position in the space-fixed frame, and a body-fixed orthogonal frame of unit vectors $(\mathbf{f}_k, \mathbf{v}_k, \mathbf{u}_k)$, where $\mathbf{u}_k$ points from bead $k$ to bead $k+1$. Rotation of $(\mathbf{f}_k, \mathbf{v}_k)$ around $\mathbf{u}_k$ is quantified by angle $\Phi_k = \alpha_k + \gamma_k$, where $(\alpha_k, \beta_k, \gamma_k)$ is the set of Euler angles which transforms $(\mathbf{f}_k, \mathbf{v}_k, \mathbf{u}_k)$ into $(\mathbf{f}_{k+1}, \mathbf{v}_{k+1}, \mathbf{u}_{k+1})$ [54]. The potential energy of the DNA chain consists of stretching, bending, torsion and electrostatic energy terms (Eq. (1) of [39]). Noteworthy, the bending rigidity was chosen so that the persistence length of the DNA chain in simulations matches the observed one, that is $L_p = 50$ nm [55], and the torsional rigidity so that the ratio of writhe to linking number difference (which depends essentially on the ratio of torsional to bending rigidity) is close to $\langle Wr \rangle / (Lk - Lk_0) \approx 0.7$ [7,8]. In simulations where the static bend of about 40° imposed by RNAP enzymes to DNA molecules [16] is taken into account, the energy term that describes the bending of the DNA chain at RNAP position $\gamma$ is modified from $\frac{1}{2} g \theta_\gamma^2$ to

$$5g(\theta_\gamma - \theta_\gamma^0)^2 \;, \tag{1}$$

where $\theta_\gamma$ denotes the angle between vectors $\mathbf{r}_{\gamma+1} - \mathbf{r}_\gamma$ and $\mathbf{r}_\gamma - \mathbf{r}_{\gamma-1}$, and $\theta_\gamma^0 = 40\pi/180$. Note that the bending force constant is increased tenfold compared to uniform DNA, in order to model the larger bending rigidity imposed by the RNAP enzyme. Electrostatic repulsion between all DNA beads $k$ and $m$ such that $|k-m| \geq 4$ (Eq. (1) of [39]) ensures that DNA strands cannot cross each other and that the linking number $Lk$ remains constant. Note that electrostatic repulsion impacts only marginally the persistence length of the DNA chain, because the Debye length (Eq. (2) of [39]) was set to 1.07 nm, which corresponds to a concentration of monovalent salts of 100 mM. The dynamics of the system was investigated by integrating numerically overdamped Langevin equations (Eqs. (3) and (4) of [39]) with a time step $\Delta t = 5$ ps.

Since each bead represents 7.5 bp and the two strands of torsionally relaxed DNA twist around the helical axis once every 10.5 bp, the linking number at equilibrium is equal to $Lk_0 = 7.5n/10.5$. Simulations were performed with linking number differences $\Delta Lk = Lk - Lk_0$ equal to $\Delta Lk = -26$ for $n = 600$ and $\Delta Lk = -130$ for $n = 2880$, corresponding to $\sigma = -0.061$ and $\sigma = -0.063$, respectively.

The fact that a motor is bound to bead $\gamma$ of the DNA chain is modeled by assuming that bead $\gamma$ behaves as a sphere of hydrodynamic radius $A = 6$ nm, which is the approximate dimension of a single RNAP [16], and by modifying accordingly the corresponding Langevin equations and body-fixed frame update rules (Eqs. (7) and (8) of [39]). Moreover, the update rules for $(\mathbf{f}_\gamma, \mathbf{v}_\gamma)$ contain an extra angular contribution $-\Omega \Delta t$, which represents the energy-consuming action of the motor [39]. For positive values of $\Omega$, this additional term increases the local density of twist for beads $k > \gamma$ ("downstream") and decreases it for beads $k < \gamma$ ("upstream"). Translocation of the motor is furthermore taken into account by incrementing $\gamma$ by $\mp 1$ each time the body-fixed frames in the neighborhood of the motor have rotated by $\pm 15\pi/10.5$. In the present work, as in [39], $\Omega$ was considered as a free parameter, which was varied over a broad range to investigate different regimes.

Finally, several sets of simulations involved a model of topological barrier, which is meant to mimic a pair of LacI repressor proteins and divides the closed DNA chain into two topologically independent loops of equal length. The action of the topological barrier is modeled by Eq. (16) of [39], which consists of two terms. The first term bridges beads $\alpha$ and $\beta$ (such that $|\beta-\alpha|=n/2$) and forces them to remain at a distance from each other of about 4 nm, which is sufficiently small to prevent other DNA segments from crossing the gap. The second term constrains the vectors of the local bases $\mathbf{f}_\alpha$ and $\mathbf{f}_\beta$ to remain nearly parallel to the vector joining beads $\alpha$ and $\beta$, thereby forbidding the rotation of $(\mathbf{f}_\alpha, \mathbf{v}_\alpha)$ and $(\mathbf{f}_\beta, \mathbf{v}_\beta)$ around $\mathbf{u}_\alpha$ and $\mathbf{u}_\beta$, respectively, and ensuring that twist cannot diffuse through beads $\alpha$ and $\beta$ [56,57].

As in [39], plectoneme dynamics was visualized by plotting the time evolution of the local writhe

$$\mathrm{Wr}(k) = \frac{1}{4\pi} \int_{k-l_w}^{k} \int_{k}^{k+l_w} \frac{(\mathbf{r}_1 - \mathbf{r}_2) \cdot (\mathrm{d}\mathbf{r}_1 \times \mathrm{d}\mathbf{r}_2)}{|\mathbf{r}_1 - \mathbf{r}_2|^3} \quad , \tag{2}$$

which is the signed counterpart of the "local unsigned writhing" discussed in [58-60] and provides a measure of the self-crossings of segment $[k-l_w, k+l_w]$. As in [39,60], $l_w$ was set to $3L_\mathrm{p}$, that is 60 beads. The apex of a plectoneme corresponds to an extremum of $\mathrm{Wr}(k)$ [58,59].

Characteristic snapshots extracted from the simulations are shown in Fig. S1 of the Supporting Information.

**RESULTS**

*Translocation speed of the motor is governed by $\Omega/n$*

Let us call $T_\gamma$ the mean torque on each side of bead $\gamma$

$$T_\gamma = \tau <(\Phi_{\gamma+2} - 2\Phi_{\gamma+1} + \Phi_\gamma)> = -\tau <(\Phi_\gamma - 2\Phi_{\gamma-1} + \Phi_{\gamma-2})> . \tag{3}$$

$\omega$, the mean rotation speed of $(\mathbf{f}_k, \mathbf{v}_k)$ around $\mathbf{u}_k$, is related to $T_\gamma$ according to [39]

$$\omega = \frac{1}{4\pi\eta a^2 l_0} \frac{2T_\gamma}{n} \quad , \tag{4}$$

where $\tau = 25\,k_\mathrm{B}T$ is the torsion force constant and $\eta = 0.00089\,\mathrm{Pa}\times\mathrm{s}$ the viscosity of the buffer. Simulations indicate that $T_\gamma$ is proportional to $\Omega$ and independent of the length $n$ of the DNA chain. $\omega$ is consequently proportional to $\Omega/n$, as is also the mean translocation speed of the motor expressed in beads per second

$$v = \frac{10.5}{7.5} \frac{\omega}{2\pi} .\qquad(5)$$

This point is illustrated in Fig. 1, which shows the evolution of $v$ with $\Omega/n$ for all simulations performed in the absence of topological barrier. In this figure, the gray dot-dashed line corresponds to

$$v = 30 \frac{\Omega}{n}\qquad(6)$$

and highlights the linear dependence of $v$ on $\Omega/n$.

### *Motors interact only weakly with the slithering motion of short DNA molecules*

As mentioned in Introduction, bacterial DNA molecules with a superhelical density of $\sigma = -0.06$ [42] form branched plectonemes, with the number of branching points increasing roughly by one every 2000 to 3000 bp [43,44]. A simple branching geometry is therefore expected for DNA chains with $n = 600$ beads (4500 bp), while the geometry of the chain may be significantly more complex for $n = 2880$ beads (21600 bp). Let us first examine how a translocating motor and the associated TSD interact with the plectoneme(s) of short chains with $n = 600$ beads.

The time evolution of the local writhe $Wr(k)$ computed from simulations with $n = 600$ and values of $\Omega/n$ ranging from 30 to 750 rad×s$^{-1}$×bead$^{-1}$ is shown in Fig. 2. For each panel, time $t$ increases along the horizontal axis from 0 to 600 ms and DNA bead index $k$ along the vertical axis from 1 to 600. The values of $Wr(k)$ computed at each time step are represented according to a color code ranging from deep blue ($Wr(k) \leq -3$) to deep red ($Wr(k) \geq 3$). The average helical density of the DNA chain being close to -0.06, strongly positive local writhe is however observed only for the largest values of $\Omega/n$ and stalled conformations, so that the color range in most panels extends from deep blue to greenish gray, which corresponds to $Wr(k) \approx 0$. The position of the motor at each time step is furthermore shown as a yellow point. Two blue lines separated by approximately $n/2 = 300$ beads and running parallel to each other are observed in each panel of Fig. 2. These blue lines correspond to minima of $Wr(k)$ and indicate the positions of the apexes of a negatively supercoiled plectoneme [58,59]. They confirm that the DNA chains form most of the time a single (unbranched) negatively supercoiled plectoneme, as expected from their length (see also the top panel of Fig. S1). Moreover, the fact that the blue lines zigzag instead of remaining horizontal reflects the slithering motion of the DNA chains, that is, their "conveyor-belt-like" motion inside the supercoil [61-64], which reverses direction randomly. A similar signature of slithering had previously been reported for a finer grained model of DNA [64]. Examination of Fig. 2 indicates that the motor has no significant effect on the slithering dynamics of the DNA chains, whatever the value of $\Omega/n$. In particular, the maximum slithering speed over long (hundreds of beads) DNA tracts is identical for all values of $\Omega/n$ and close to 4000 beads per second. In each panel, a sequence when the DNA chain slithers at the maximum speed is highlighted with a brown dashed line.

At this point, a few words may be useful to clarify why, in Fig. 2, the yellow lines showing the position of the motor are not fully straight. As described in Theoretical Methods, the motor induces at each time step an infinitesimal rotation of angle $-\Omega\Delta t$ of the local basis at bead $\gamma$ and acts consequently as a constant torque motor. It would also act as a constant speed motor (and the yellow lines in Fig. 2 would be fully straight) in the absence of torsional fluctuations along the DNA chain. However, torsional fluctuations transiently increase or decrease the rotational speed of the local basis at bead $\gamma$ and consequently let the translational speed of the motor fluctuate around the average value imposed by $\Omega$. This, in turn, makes the yellow lines in Fig. 2 somewhat wavy, especially for low values of $\Omega$.

Further information on plectoneme dynamics is gained from the blue solid lines in Fig. 3, which show the probability $Q(\Delta k)$ that the local writhe at bead $\gamma+\Delta k$ satisfies $Wr(\gamma+\Delta k) \leq -1.4$ for increasing values of $\Omega/n$. Function $Q(\Delta k)$ is rescaled so that $Q(\Delta k)=1$ for a uniform distribution. Values of the local writhe smaller than -1.4 are observed only at the apex of a plectoneme, so that $Q(\Delta k)$ actually indicates whether the probability of finding the apex of a plectoneme at position $\Delta k$ from the motor is larger ($Q(\Delta k)>1$) or smaller ($Q(\Delta k)<1$) than average. Up to $\Omega/n = 150$ rad×s$^{-1}$×bead$^{-1}$, the plots of $Q(\Delta k)$ do not differ significantly from a uniform distribution, whereas for $\Omega/n \geq 300$ rad×s$^{-1}$×bead$^{-1}$ a marked hole is observed for small positive values of $\Delta k$, that is immediately downstream of the motor. This hole reflects the destabilization of negative supercoils due to the motor injecting positive twist downstream of its position.

We next asked ourselves whether the static bend of about 40° imposed by RNAP enzymes to DNA molecules [16] is able to affect significantly plectoneme dynamics and, in particular, to pin the apexes of the plectonoemes [21,48,49]. In order to answer this question, the energy term that describes the bending of the DNA chain at position $\gamma$ was modified according to Eq. (1). The time evolution of $Wr(k)$ computed from simulations with the modified bending term is shown in Fig. 4 for values of $\Omega/n$ ranging from 30 to 750 rad×s$^{-1}$×bead$^{-1}$. One observes that the trajectories of the motor (yellow lines) are indeed somewhat more prone to superpose with the trajectories of the apexes of the plectonemes (blue lines) in Fig. 4 than in Fig. 2, especially for the lowest values of $\Omega/n$. This observation is confirmed by the red dashed lines in Fig. 3, which show the evolution of $Q(\Delta k)$ when the static bend at $\gamma$ is taken into account. However, Fig. 3 also indicates that the increase in the probability of finding the motor close to the apex of the plectoneme remains rather modest (maximum $Q(0) \approx 2.3$ for $\Omega/n = 30$ rad×s$^{-1}$×bead$^{-1}$). Moreover, the static bend at $\gamma$ alters only slightly the maximum slithering speed. Indeed, the brown dashed lines in Fig. 4 highlight sequences when the DNA chain slithers at the maximum speed of 5600 beads per second, that is, only slightly faster than the maximum speed of 4000 beads per second observed in Fig. 2. It is not even clear whether this difference is statistically meaningful or not.

We finally checked the predictions of the model when a topological barrier bridges beads $\alpha = 1$ and $\beta = 301$ and separates the DNA chain into two independent loops of equal length (see Methods). The motor was initially positioned at bead $\gamma = 451$, that is, at the center of one loop, and simulations were performed using the same protocol as in the absence of topological barrier. The time evolution of the local writhe $Wr(k)$ is shown in Fig. S2 of the Supporting Information for simulations with a static bend of 40° at bead $\gamma$ and values of $\Omega/n$ ranging from 30 to 750 rad×s$^{-1}$×bead$^{-1}$. The horizontal brown dashed line in each panel indicates the position of bead $\beta = 301$. Because beads $\alpha$ and $\beta$ are constrained to remain at a short distance from one another by the topological barrier, the slithering motion cannot set in, which is reflected in the blue lines (showing the position of the apexes of the plectoneme) remaining parallel to the horizontal axis and at equal distances from the brown line. More surprising is the fact that the motor (yellow lines) diffuses a lot in all panels, except for $\Omega/n = 30$ rad×s$^{-1}$×bead$^{-1}$, although it is expected to stall after a short while, because of the build-up of a resisting torque downstream of its position [39]. Extensive backward motion is even observed, although the "active" angular contribution $-\Omega\Delta t$ induces forward translocation (see Methods). In real life, RNAP enzymes are admittedly able to backtrack along the DNA template for distances as long as the length of the RNA transcript [65], but this is in general due to abortive transcription initiation, regulatory DNA sequences, or errors in the growing RNA transcript [65-68], which are not taken into account in the present model. The explanation for this unexpected result is presumably that the system schematically consists of one heavy bead (the motor) and one relatively rigid plectoneme, which rotate with respect to one another. This relative rotation modifies the twist angles in the neighborhood of the motor and modifies its position along the DNA chain according to the translocation rules sketched in Methods. Diffusion of the motor in Fig. S2 is therefore not relevant but results essentially from the limitations of the model and, first of all, from the fact that in the model the rotation of the body-fixed frames triggers the translocation of the motor, whereas in real life it is instead the translocation of the motor that unfolds the double helix. We note that the forward translocation of the motor in all panels of Figs. 2 and 4 is significantly faster than its diffusion in Fig. S2, so that these limitations could safely be ignored in the preceding paragraphs.

*Motors stimulate the diffusion and growth/shrinkage modes of the plectonemes of long DNA molecules*

Let us now examine how a translocating motor and the associated TSD affect the plectonemes of longer DNA chains with $n = 2880$ beads (21600 bp), whose branching geometry is expected to be significantly more complex than for $n = 600$ beads (4500 bp). The corresponding plots of $\text{Wr}(k)$ are shown in Fig. 5 (for simulations where the static bend at $\gamma$ is disregarded) and Fig. 6 (for simulations where the static bend at $\gamma$ is taken into account). These figures indicate that DNA chains are usually composed of 6 to 8 plectonemes up to $\Omega/n = 150$ rad×s$^{-1}$×bead$^{-1}$, and 4 or 5 plectonemes for $\Omega/n \geq 300$ rad×s$^{-1}$×bead$^{-1}$. It is moreover important to realize that these plectonemes are not isolated along the DNA chain, but that each plectoneme has instead two neighbors with whom it shares a common foot, so that the plectonemes altogether occupy the whole DNA chain. This point is illustrated in Figs. S3 and S4 of the Supporting Information, which show representative contact maps extracted from the simulations in Figs. 5 and 6, respectively. In these contact maps, computed as described in [39], plectonemes appear as brown segments parallel to the secondary diagonal. One clearly sees that a plectoneme starts where the previous one ends and that all DNA beads belong to a plectoneme. The immediate consequence is that the dynamics of a given plectoneme is necessary coupled to the dynamics of its neighbors. Practically, synchronous slithering (all plectonemes slither at the same speed in the same direction), which corresponds to the case where all blue lines in Figs. 5 and 6 would run parallel to each other, is never observed. Rather, the apex of each plectoneme may approach closer or separate further from the apex of neighboring ones. Since these plectonemes have common feet, this implies that the diffusion of plectonemes is in most cases accompanied by some variation of their own length and/or the length of their neighbors.

Most striking in Figs. 5 and 6 is the fact that, up to $\Omega/n = 150$ rad×s$^{-1}$×bead$^{-1}$, the yellow line systematically runs for several tens of ms either on top of a blue line or nearly parallel to the blue line located immediately below. As can be checked in Figs. S3 and S4, this indicates that the motor sits either at the apex of a plectoneme or on its downstream arm and that the translocation of the motor and the slithering/translation of the DNA chain are able to synchronize for a while, as if the motor would trail behind itself the plectoneme on which it sits. Forward translocation of the motor and the synchronized slithering/translation of the plectoneme are furthermore accompanied by a significant increase in the length of the plectoneme. This point is illustrated in Fig. 7, which shows superposed contact maps extracted at $t = 15$ ms (red) and $t = 35$ ms (blue) from the simulation with $\Omega/n = 150$ rad×s$^{-1}$×bead$^{-1}$ shown in Fig. 6 (the conformation of the DNA chain at $t = 15$ ms is also shown in the bottom panel of Fig. S1). In Fig. 7, the red and blue dot-dashed lines show the respective locations of the motor. During the 20 ms interval, the motor moves forward by 105 beads, while the apex of the plectoneme on which it sits (labeled A in Fig. 7 and shown as a blue tube in Fig. S1) and that of the plectoneme located immediately downstream (labeled B in Fig. 7 and shown as a green tube in Fig. S1) move in the same direction by 80 and 90 beads, respectively. The length of plectoneme A simultaneously increases by about 220 beads, while that of plectoneme B decreases by the same amount. In contrast, the plectoneme located upstream of the motor (labeled C) does not move and its length remains constant. The TSD generated by the motor during its translocation over a certain length is consequently responsible for the slithering/translation of plectonemes A and B by approximately the same length, as well as the lengthening of plectoneme A and shortening of plectoneme B by twice this length. The DNA tracts made available by the forward motion of plectoneme A and the reduction in length of plectoneme B wind around each other and add turns to plectoneme A.

At this point, it is worth remembering that single-molecule experiments have shown that the spontaneous displacements of plectonemes at moderate superhelical density consist of diffusion events and growth/shrinkage events, the latter ones corresponding to the case where a plectoneme at a certain position shrinks (and eventually "terminates" (disappears)) while a plectoneme at another, possibly distant position simultaneously grows (eventually after "nucleating" (building from scratch)) [21,69,70]. As discussed in the preceding paragraph, the dynamics displayed in Fig. 7 can precisely be decomposed into diffusion and growth/shrinkage events. Conclusion is therefore that the TSD generated by the translocating motor interacts with the plectonemes of long DNA molecules by stimulating their spontaneous diffusion and growth/shrinkage modes, however in a coordinated and oriented manner, which results in the forward translation and growth of plectoneme A.

It is noteworthy that the motor translocation speed of 105 beads in 20ms and the plectoneme translation speed of 80 or 90 beads in 20 ms observed in Fig. 7 are close to the maximum slithering speed of 4000 or 5600 beads per second observed in the simulations with short DNA chains (Figs. 2 and 4). For values of $\Omega/n$ larger than 300 rad×s$^{-1}$×bead$^{-1}$, the translocation speed of the motor becomes significantly larger than the spontaneous (thermal) slithering/translation of plectonemes. As a consequence, the motor would have to somehow inject significant amounts of energy in the plectoneme on which it sits in order to allow for synchronous motion of the motor and the plectoneme. The motor is clearly not able to do that, as can be deduced from the fact that for $\Omega/n \geq 300$ rad×s$^{-1}$×bead$^{-1}$ the blue and yellow lines in Figs. 5 and 6 seldom run parallel to each other. Instead, quasi immobile plectonemes nucleate almost periodically at the upstream side of the motor, grow up to several kbp, detach from the motor, shrink and disappear. A detailed analysis of the simulations indicates that the TSD still stimulates the spontaneous diffusion and growth/shrinkage modes of neighboring plectonemes, but the details of the dynamics are rather different than for lower speeds.

A typical example is illustrated in Fig. 8, which shows superposed contact maps extracted at $t = 51$ ms (red) and $t = 71$ ms (blue) from the simulation with $\Omega/n = 300$ rad×s$^{-1}$×bead$^{-1}$ shown in Fig. 6. The motor sits at the common foot of plectonemes A (located upstream) and B (located downstream). During the 20 ms interval, the motor moves forward by 172 beads. The apex of plectonemes A and B do not move during the translocation of the motor, but plectoneme B transfers about 140 beads to plectoneme A via the growth/shrinkage mode. Simultaneously, the wave of negative supercoiling traveling upstream of the motor lets plectoneme C diffuse backward by about 140 beads. These beads wind around those originating from plectoneme B, thereby adding turns to plectoneme A. The consequence is that the length of plectoneme A increases by about 280 beads in the 20 ms interval. The fact that the motor still sits on the downstream arm of a plectoneme, but preferentially at its downstream foot (or close to it) rather than at the apex, is a feature which is common to all simulations performed with large values of $\Omega/n$, as can be checked in the right panels of Figs. S3 and S4. The backward diffusion of the plectoneme located upstream of the motor is also a very general feature. In the right panels of Figs 5 and 6, this backward diffusion causes the blue lines located immediately below the yellow lines to bend downward as soon as new blue lines emerge on top of the yellow lines.

Two additional pieces of information are provided by Fig. 9, which shows the evolution of $Q(\Delta k)$ for DNA chains with $n = 2880$ beads when the static bend at bead $\gamma$ is disregarded (blue solid lines) or taken into account (red dashed lines). The first conclusion that can be drawn from this figure is that, for long DNA chains, the static bend at $\gamma$ does not increase significantly the probability that the RNAP sits at the apex of the plectoneme. The second information is that, for $\Omega/n \geq 750$ rad×s$^{-1}$×bead$^{-1}$, the perturbations due to the motor extend over all the DNA chain, that is over more than 20000 bp, in agreement with recent experiments which have shown that supercoiled domains span up to 25 kbp on both sides of a transcribing RNAP [22].

We finally checked the predictions of the model when the DNA chain is separated into two independent loops of equal length by a topological barrier which bridges beads $\alpha = 1$ and $\beta = 1441$ (see Methods). The motor was initially positioned at $\gamma = 2161$, that is, at the center of one loop, and simulations were performed using the same protocol as in the absence of topological barrier. The time evolution of $Wr(k)$ computed from these simulations is shown in Fig. S5 of the Supporting Information for the model with a static bend of 40° at position $\gamma$ and values of $\Omega/n$ ranging from 30 to 1500 rad×s$^{-1}$×bead$^{-1}$. The horizontal brown dashed line in each panel indicates the position of bead $\beta = 1441$. Examination of Fig. S5 suggests that for all values of $\Omega/n$ the system reaches a steady state within a few tens of ms and that the relative rotation of the motor bead and the rest of the DNA coil, which plagued the simulations with $n = 600$ beads, does not affect significantly the results for $n = 2880$ beads, at least in the investigated time interval. The reason is probably that, for $n = 2880$, the DNA chain forms a branched plectoneme, whose rotational dynamics is much slower than that of a single plectoneme composed of 600 beads. The characteristic time scale of the relative motion of the motor bead and the rest of the DNA coil becomes consequently sufficiently long to leave the results shown in Fig. S5 unperturbed. Most importantly, Fig. 10, which displays contact maps extracted at $t = 50$ ms from the simulations shown in Fig. S5, indicates that, up to $\Omega/n = 300$ rad×s$^{-1}$×bead$^{-1}$, the loop which contains the motor reorganizes so that the motor sits at the apex of a long negatively supercoiled plectoneme. In contrast, for $\Omega/n \geq 750$ rad×s$^{-1}$×bead$^{-1}$, the motor sits at the foot of the plectoneme, in a region where $Wr(k)$ is markedly positive. Fig. 10 actually compares well with Fig. S4, which shows representative contact maps extracted from simulations without the topological barrier and also displays a marked displacement of the motor from the apex to the foot of the plectoneme upon increase of $\Omega/n$. Conclusion is therefore that the localization of the stalled motor in a system that contains topological barriers reflects its preferential localization in the unconstrained system.

**DISCUSSION**

In this paper, we performed Brownian Dynamics simulations with a coarse-grained model to understand how the waves of positive and negative supercoiling generated by a motor which translocates along a bacterial DNA molecule influences the dynamics of its plectonemes. These simulations provide clear answers to some of the questions that were raised in Introduction. More precisely,

- the length of the DNA molecule does matter: Short molecules up to a few thousand base pairs form a single plectoneme [43,44], whose slithering dynamics is only marginally affected by the activity of a motor. In contrast, motors impact strongly the dynamics of the plectonemes of longer DNA molecules, which form branched plectonemes [43,44]. Quite interestingly these interactions proceed via the stimulation of the spontaneous (thermal) diffusion and growth/shrinkage displacement modes of the plectonemes [21,69,70] and dissipate therefore little energy. This is possibly the most significant result of the present work.

- simulations suggest that motor/plectonemes interactions are only marginally affected by the static bend that the motor eventually imposes to the DNA molecule [16,50]. The reason is probably that the pinning energy associated with a static bend is significantly weaker than the energy dissipated by the TSD.

- according to the simulations, the localization of the motor in a stalled system that contains topological barriers reflects the preferential localization of the motor in the unconstrained system. While the motor always locates on the downstream arm of a plectoneme, it sits preferentially close to the apex at low exerted torque, but close to the foot at large torque. Localization of the RNAP enzyme in stalled conformations could eventually be investigated experimentally using single-molecule set-ups inspired by [21], where the RNAP was labeled with Alexa647 and the system containing also Sytox Orange intercalating dye was visualized through Dual-color HiLo fluorescence microscopy.

- motor/plectonemes interactions in long DNA molecules depend crucially on whether the motor translocates more slowly or more rapidly than the maximum slithering/translation speed of plectonemes. If translocation is slower than slithering/translation, then the slithering of the plectoneme on which the motor sits and that of the plectoneme located downstream of the motor synchronize with the motion of the motor. In contrast, if translocation is faster than slithering/translation, then the plectoneme on which the motor sits remains almost immobile during its growth and the plectoneme located upstream diffuses backward.

Concerning this last point, the model itself provides no indication regarding the relative speeds of motor translocation and plectoneme slithering/translation. Indeed, in the model, the translocation speed of the motor is governed by $\Omega/n$ (see Fig. 1), which was considered as a free parameter and varied over a broad range to investigate different regimes. From the experimental point of view, the translocation speed of an RNAP enzyme (about 50 bp×s$^{-1}$ [45-47]) and a replisome (about 500 bp×s$^{-1}$ [26,27]) are well-known. In contrast, the speed of plectoneme slithering/translation is, to the best of our knowledge, not known precisely. *In vitro* experiments indicate that plectonemes can slither/translate at speeds up to about 6000 bp×s$^{-1}$ for a modest stretching force of 0.8 pN (Figs. 2B and 2D of [69]). However, the associated diffusion coefficient is divided by about 5 for a stretching force of 1.6 pN and by more than 20 for a stretching force of 3.2 pN (Fig. 2F of [69]). The point is that both RNAP enzymes and replisomes exert a significant tension on DNA molecules. Since RNAP enzymes can work against resisting forces up to 25 pN [13] and replisomes against resisting forces up to 34 pN [14], it is probable that tension in the DNA template during transcription or replication reaches several pN, or more, and that the slithering/translation speed of plectonemes *in vivo* is much lower than the 6000 bp×s$^{-1}$ observed in [69]. A reasonable guess is that the slithering/translation speed of plectonemes is larger than transcription speed and smaller than replication speed, but this point remains to be ascertained.

Finally, it should be mentioned that for DNA tension larger than about 0.5 pN, underwound or torsionally relaxed B-DNA may partially convert to left-handed DNA forms [71,72]. In particular, transcription is known to be associated with Z-DNA formation [73,74]. Z-DNA tracts form upstream of the RNAP, where they relieve part of the negative torsional stress. This in turn tends to homogenize superhelical density on both sides of the RNAP and may consequently mitigate to some extent the effects discussed above. The impact is however probably rather limited, because Z-DNA is unstable and reverts rapidly to B-DNA.

**CONCLUSION**

A coarse-grained model and Brownian Dynamics simulations were used to understand how the TSD generated by a translocating RNAP enzyme or a translocating replisome interacts with the plectonemes of bacterial DNA. Simulations revealed that the TSD couples efficiently the dynamics of the motor and the plectonemes of long DNA molecules by stimulating in a coordinated manner the spontaneous displacement modes of the plectonemes. For the sake of a clear understanding of the coupled dynamics, the model was kept as simple as possible and consists exclusively of a circular DNA chain, a motor modeled as an effective external torque, and eventually a topological barrier modeled as a set of constraints between two DNA beads. This model can be improved along several lines. First, in the present version of the model, the rotation of the body-fixed frames triggers the translocation of the motor, whereas in real life it is instead the translocation of the motor that unfolds the double helix. While it is not expected that implementing a constant translation speed motor instead of a constant torque one would impact significantly the results discussed above, it would however represent a preliminary step for further improvements. For example, there is currently a debate concerning the role of the length of RNA transcripts. Indeed, early experiments suggested that a long RNA transcript is needed for the generation of the TSD, the proposed explanation being that an RNAP enzyme bound to a short RNA transcript merely rotates around the DNA molecule during translocation, whereas an RNAP enzyme with a long RNA transcript experiences a larger rotational drag and forces instead the DNA molecule to rotate around its helical axis [19]. In contrast, very recent single-molecule experiments suggest instead that induction of DNA supercoiling by transcription is independent of the RNA transcript [21]. It would consequently be interesting to introduce the growing RNA transcript in the coarse-grained model by adding one bead to a linear RNA chain each time the motor translocates by one bead along the circular DNA chain and check whether Brownian Dynamics simulations can shed some light on its potential role. The increasing interactions between the RNA transcript and the DNA chain, as well as the increasing hydrodynamic drag of the RNA transcript, will certainly impact significantly the relationship between the translational and the rotational speeds of the motor, but it is expected that simulations will also reveal additional and subtler features of this intricate mechanism. Moreover, the motor stalls very rapidly in simulations involving a topological barrier, because of the building of a resisting torque downstream of the motor. In real life, this problem is solved by the recruitment of Type I topoisomerase enzymes upstream of the motor and Type II topoisomerase enzymes (gyrases) downstream of the motor. These enzymes break temporarily one (respectively, two) DNA strands and relax negative (respectively, positive) supercoils [28,75-77]. Introduction of topoisomerases in the coarse-grained model would give access to a more complete understanding of the dynamics of DNA replication and transcription [35,78].

**SUPPORTING INFORMATION**

Figures S1 to S5, showing representative snapshots extracted from the simulations, additional plots of the local writhe $Wr(k)$, and additional contact maps (PDF).

**ACKNOWLEDGMENT**

This work was supported by Centre National de la Recherche Scientifique and Université Grenoble Alpes. It did not receive any specific grant from funding agencies in the public, commercial, or not-for-profit sectors.

**REFERENCES**

1. White, J.H. 1969. Self-linking and the Gauss integral in higher dimensions. *Am. J. Math.* 91:693–728.

2. Cãlugãreanu, G. 1961. Sur les classes d'isotopie des noeuds tridimensionnels et leurs invariants. *Czech. Math. J.* 11: 588-625.

3. Fuller, F.B. 1968. The writhing number of a space curve. *Proc. Natl. Acad. Sci. U.S.A.* 68:815–819.

4. Dennis, M.R., and J.H. Hannay. 2005. Geometry of Calugareanu's theorem. Proc. R. Soc. A 461, 3245–3254.

5. Fuller, F.B. 1978. Decomposition of the linking number of a closed ribbon: A problem from molecular biology. *Proc. Natl. Acad. Sci. U.S.A.* 75:3557–3561.

6. Bates, A.D., and A. Maxwell. 2005. DNA topology. Oxford University Press, New York.

7. Boles, T.C., J.H. White, and N.R. Cozzarelli. 1990. Structure of plectonemically supercoiled DNA. *J. Mol. Biol.* 213:931-951.

8. Bednar, J., P. Furrer, A. Stasiak, J. Dubochet, E.H. Egelman, and A.D. Bates. 1994. The twist, writhe and overall shape of supercoiled DNA change during counterion-induced transition from a loosely to a tightly interwound superhelix. *J. Mol. Biol.* 235: 825–847.

9. Ishihama, A. 2000. Functional modulation of Escherichia coli RNA polymerase. *Annu. Rev. Microbiol.* 54:499-518.

10. Weng, X., C.H. Bohrer, K. Bettridge, A.C. Lagda, C. Cagliero, D.J. Jin, and J. Xiao. 2019. Spatial organization of RNA polymerase and its relationship with transcription in Escherichia coli. *Proc. Natl. Acad. Sci. USA.* 116:20115-20123.

11. Schafer, D.A., J. Gellest, M.P. Sheetz, and R. Landick. 1991. Transcription by single molecules of RNA polymerase observed by light microscopy. *Nature.* 352:444-448.

12. Yin, H., R. Landick, and J. Gelles. 1994. Tethered Particle Motion method for studying transcript elongation by a single RNA polymerase molecule. *Biophys. J.* 67:2468-2478.

13. Wang, M.D., M.J. Schnitzer, H. Yin, R. Landick, J. Gelles, and S.M. Block. 1998. Force and velocity measured for single molecules of RNA polymerase. *Science*. 282:902-907.

14. Wuite, G.J.L., S.B. Smith, M. Young, D. Keller, and C. Bustamante. 2000. Single molecule studies of the effect of template tension on T7 DNA polymerase activity. *Nature*. 404:103-106.

15. Thomen, P., P.J. Lopez, U. Bockelmann, J. Guillerez, M. Dreyfus, and F. Heslot. 2008. T7 RNA polymerase studied by force measurements varying cofactor concentration. *Biophys. J.* 95:2423–2433.

16. Yin, Y.W., and T.A. Steitz. 2002. Structural basis for the transition from initiation to elongation transcription in T7 RNA polymerase. *Science*. 298:1387-1395.

17. Liu, L., and J.C. Wang. 1987. Supercoiling of the DNA template during transcription. *Proc. Natl. Acad. Sci. USA*. 84:7024-7027.

18. Wu, H.Y., S. Shyy, J.C. Wang, and L.F. Liu. 1988. Transcription generates positively and negatively supercoiled domains in the template. *Cell*. 53:433-440.

19. Tsao, Y.P., H.Y. Wu, and L.F. Liu. 1989. Transcription-driven supercoiling of DNA: Direct biochemical evidence from in vitro studies. *Cell*. 56:111-118.

20. Kim, S., B. Beltran, I. Irnov, and C. Jacobs-Wagner. 2019. Long-distance cooperative and antagonistic RNA polymerase dynamics via DNA supercoiling. *Cell*. 179:106–119.

21. Janissen, R., R. Barth, M. Polinder, J. van der Torre, and C. Dekker. 2024. Single-molecule visualization of twin-supercoiled domains generated during transcription. *Nucleic Acids Res.* 52:1677–1687.

22. Visser, B.J., S. Sharma, P.J. Chen, A.B. McMullin, M.L. Bates, and D. Bates. 2022. Psoralen mapping reveals a bacterial genome supercoiling landscape dominated by transcription. *Nucleic Acids Res.* 50:4436-4449.

23. Ma, J., L. Bai, and M.D. Wang. 2013. Transcription under torsion. *Science*. 40:1580-1583.

24. Xu, Z.Q., and N.E. Dixon. 2018. Bacterial replisomes. *Curr. Opin. Struct. Biol.* 53:159–168.

25. Graham, J.E., K.J. Marians, and S.C. Kowalczykowski. 2017. Independent and stochastic action of DNA polymerases in the replisome. *Cell*. 169:1201-1213.

26. Reyes-Lamothe, R., C. Possoz, O. Danilova, and D.J. Sherratt. 2008. Independent positioning and action of Escherichia coli replisomes in live cells. *Cell*. 133:90–102.

27. Bhat, D., S. Hauf, C. Plessy, Y. Yokobayashi, S. Pigolotti. 2022. Speed variations of bacterial replisomes. *eLife*. 11:e75884.

28. Ma, J., and M.D. Wang. 2016. DNA supercoiling during transcription. *Biophys. Rev.* 8:S75-S87.

29. Brackley, C.A., J. Johnson, A. Bentivoglio, S. Corless, N. Gilbert, G. Gonnella, and D. Marenduzzo. 2016. Stochastic model of supercoiling-dependent transcription. *Phys. Rev. Lett.* 117:018101.

30. Meyer, S., and G. Beslon. 2014. Torsion-mediated interaction between adjacent genes. *PLoS Comput. Biol.* 10:e1003785.

31. Tripathi, S., S. Brahmachari, J.N. Onuchic, and H. Levine. 2022. DNA supercoiling-mediated collective behaviour of co-transcribing RNA polymerases. *Nucleic Acids Res.* 50:1269-1279.

32. Hatfield, G.W., and C.J. Benham. 2002. DNA topology-mediated control of global gene expression in Escherichia coli. *Annu. Rev. Genet.* 36:175-203.

33. Racko, D., F. Benedetti, J. Dorier, Y. Burnier, and A. Stasiak. 2015. Generation of supercoils in nicked and gapped DNA drives DNA unknotting and postreplicative decatenation. *Nucleic Acids Res.* 43:7229-7236.

34. Benedetti, F., D. Racko, J. Dorier, Y. Burnier, and A. Stasiak. 2017. Transcription-induced supercoiking explains formation of self-interacting chromatin domains in S. pombe. *Nucleic Acids Res.* 45:9850-9859.

35. Racko, D., F. Benedetti, J. Dorier, and A. Stasiak. 2018. Transcription-induced supercoiling as the driving force of chromatin loop extrusion during formation of TADs in interphase chromosomes. *Nucleic Acids Res.* 46:1648-1660.

36. Ruskova, R., and D. Racko. 2021. Entropic competition between supercoiled and torsionally relaxed chromatin fibers drives loop extrusion through pseudo-topologically bound cohesin. *Biology*. 10:130.

37. Sevier, S.A., and H. Levine. 2017. Mechanical properties of transcription. *Phys. Rev. Lett.* 118: 268101.

38. Fosado, Y.A.G., D. Michieletto, C.A. Brackley, and D. Marenduzzo. 2021. Nonequilibrium dynamics and action at a distance in transcriptionally driven DNA supercoiling. *Proc. Natl. Acad. Sci. USA*. 118:e1905215118.

39. Joyeux, M. 2024. Transcribing RNA polymerases: Dynamics of twin supercoiled domains. *Biophys. J.* 123:3898-3910.

40. Mielke, S.P., W.H. Fink, V.V. Krishnan, N. Grønbech-Jensen, and C.J. Benham. 2004. Transcription-driven twin supercoiling of a DNA loop: A Brownian dynamics study. *J. Chem. Phys.* 121:8104-8112.

41. Racko, D., F. Benedetti, J. Dorier, Y. Burnier, and A. Stasiak. 2017. Molecular dynamics simulation of supercoiled, knotted, and catenated DNA molecules, including modeling of action of DNA gyrase. *Methods Mol. Biol.* 1624:339-372.

42. Bauer, W.R. 1978. Structure and reactions of closed duplex DNA. *Annu. Rev. Biophys. Bioeng.* 7:287-313.

43. Boles, T.C., J.H. White, and N.R. Cozzarelli. 1990. Structure of plectonemically supercoiled DNA. *J. Mol. Biol.* 213:931-951.

44. Hacker, W.C., S. Li, and A.H. Elcock. 2017. Features of genomic organization in a nucleotide-resolution molecular model of the Escherichia coli chromosome. *Nucleic Acids Res.* 45:7541–7554.

45. Chong, S., C. Chen, H. Ge, and X.S. Xie. 2014. Mechanism of transcriptional bursting in bacteria. *Cell*. 158:314-325.

46. Rovinskiy, N., A.A. Agbleke, O. Chesnokova, Z. Pang, and N.P. Higgins. 2012. Rates of gyrase supercoiling and transcription elongation control supercoil density in a bacterial chromosome. *PLoS Genet*. 8: e1002845.

47. Vogel, U., and K.F. Jensen. 1994. The RNA chain elongation rate in *Escherichia coli* depends on the growth rate. *J. Bacteriol.* 176:2807-2813.

48. Kim, S.H., M. Ganji, E. Kim, J. van der Torre, E. Abbondanzieri, and C. Dekker. 2018. DNA sequence encodes the position of DNA supercoils. *eLife*. 7:e36557.

49. Desai, P.R., S. Brahmachari, J.F. Marko, S. Das, and K.C. Neuman. 2020. Coarse-grained modelling of DNA plectoneme pinning in the presence of base-pair mismatches. *Nucleic Acids Res.* 48:10713–10725.

50. Sreenivasan, R., S. Heitkamp, M. Chhabra, R. Saecker, E. Lingeman, M. Poulos, D. McCaslin, M.W. Capp, I. Artsimovitch, and M.T. Record,M.T. 2016. Fluorescence resonance energy transfer characterization of DNA wrapping in closed and open Escherichia coli RNA polymerase−λP$_R$ promoter complexes. *Biochemistry*. 55:2174–2186.

51. Leng, F., B. Chen, and D.D. Dunlap. 2011. Dividing a supercoiled DNA molecule into two independent topological domains. *Proc. Natl. Acad. Sci. USA*. 108:19973-19978.

52. Leng, F., L. Amado, and R. McMacken. 2004. Coupling DNA supercoiling to transcription in defined protein systems. *J. Biol. Chem.* 279:47564–47571.

53. Dorman, C.J. 2019. DNA supercoiling and transcription in bacteria: a two-way street. *BMC Mol. Cell Biol.* 20:26.

54. Chirico, G., and J. Langowski. 1994. Kinetics of DNA supercoiling studied by Brownian dynamics simulation. *Biopolymers*. 34:415-433.

55. Baumann, C.G., S.B. Smith, V.A. Bloomfield, and C. Bustamante. 1997. Ionic effects on the elasticity of single DNA molecules, *Proc. Natl. Acad. Sci. USA*. 94:6185–6190.

56. Joyeux, M., and I. Junier. 2020. Requirements for DNA-bridging proteins to act as topological barriers of the bacterial genome. *Biophys. J.* 119:1215-1225.

57. Joyeux, M. 2022. Models of topological barriers and molecular motors of bacterial DNA. *Mol. Simul.* 48:1688-1696.

58. Vologodskii, A.V., S.D. Levene, K.V. Klenin, M. Frank-Kamenetskii, and N.R. Cozzarelli. 1992. Conformational and thermodynamic properties of supercoiled DNA. *J. Mol. Biol.* 227:1224-1243.

59. Klenin, K., and J. Langowski. 2000. Computation of writhe in modeling of supercoiled DNA. *Biopolymers*. 54:307–317.

60. Michieletto, D. 2016. On the tree-like structure of rings in dense solutions. *Soft Matter*. 12:9485-9500.

61. Marko, J.F. 1997. The internal 'slithering' dynamics of supercoiled DNA. *Physica A*. 244: 263-277.

62. Brouns, T., H. de Keersmaecker, S.F. Konrad, N. Kodera, T. Ando, J. Lipfert, S. de Feyter, and W. Vanderlinden. 2018. Free energy landscape and dynamics of supercoiled DNA by high-speed atomic force microscopy. *ACS Nano*. 12:11907−11916.

63. Huang, J., T. Schlick, and A. Vologodskii. 2001. Dynamics of site juxtaposition in supercoiled DNA. *Proc. Natl. Acad. Sci. USA*. 98: 968–973.

64. Sprous, D., and S.C. Harvey. 1996. Action at a distance in supercoiled DNA: Effects of sequence on slither, branching, and intramolecular concentration. *Biophys. J.* 70:1893-1908.

65. Komissarova, N., and M. Kashlev. 1997. Transcriptional arrest: Escherichia coli RNA polymerase translocates backward, leaving the 3' end of the RNA intact and extruded. *Proc. Natl. Acad. Sci. USA*. 94:1755–1760.

66. Artsimovitch, I., and R. Landick. 2000. Pausing by bacterial RNA polymerase is mediated by mechanistically distinct classes of signals. *Proc. Natl. Acad. Sci. USA*. 97:7090–7095.

67. Lerner, E., S.Y. Chung, B.L. Allen, S. Wang, J. Lee, S.W. Lu, L.W. Grimaud, A. Ingargiola, X. Michalet, Y. Alhadid, *et al*. 2016. Backtracked and paused transcription initiation intermediate of Escherichia coli RNA polymerase. *Proc. Natl. Acad. Sci. USA*. 113:E6562–E6571.

68. Abdelkareem, M., C. Saint-André, M. Takacs, G. Papai, C. Crucifix, X. Guo, J. Ortiz, and A. Weixlbaumer. 2019. Structural basis of transcription: RNA polymerase backtracking and its reactivation. *Mol. Cell*. 75:298–309.

69. van Loenhout, M.T.J., M.V. de Grunt, and C. Dekker. 2012. Dynamics of DNA supercoils. *Science*. 338:94-97.

70.     Ganji, M., S.H. Kim, J. van der Torre, E. Abbondanzieri, and C. Dekker. 2016. Intercalation-based single-molecule fluorescence assay to study DNA supercoil dynamics. *Nano Lett*. 16:4699–4707.

71.     Marko, J.F. and S. Neukirch. 2013. Global force-torque phase diagram for the DNA double helix: structural transitions, triple points, and collapsed plectonemes. *Phys. Rev. E.* 88:062722.

72.     Yan, Y., Y. Ding, F. Leng, D. Dunlap, and L. Finzi. 2018. Protein-mediated loops in supercoiled DNA create large topological domains. *Nucleic Acids Res.* 46:4417-4424.

73.     Wittig, B., T. Dorbic, and A. Rich. 1991. Transcription is associated with Z-DNA formation in metabolically active permeabilized mammalian cell nuclei. *Proc. Natl. Acad. Sci. USA.* 88:2259–2263.

74.     Wolfl, S., C. Martinez, A. Rich, and J. A. Majzoub. 1996. Transcription of the human corticotropin-releasing hormone gene in NPLC cells is correlated with Z-DNA formation. *Proc. Natl. Acad. Sci. USA*. **93:**3664–3668.

75.     Reece, R.J., and A. Maxwell. 1991. DNA gyrase: structure and function. *Crit. Rev. Biochem. Mol. Biol.* 6:335–375.

76.     McKie, S.J., K.C. Neuman, and A Maxwell. 2021. DNA topoisomerases: Advances in understanding of cellular roles and multi-protein complexes via structure-function analysis. *BioEssays*. 43:2000286.

77.     Ahmed, W., C. Sala, S.R. Hegde, R.Kumar Jha, S.T. Cole, and V. Nagaraja. 2017. Transcription facilitated genome-wide recruitment of topoisomerase I and DNA gyrase. *PLoS Genet.* 13:e1006754.

78.     Baranello, L., D. Wojtowicz, K. Cui, B.N. Devaiah, H.J. Chung, K.Y. Chan-Salis, R. Guha, K. Wilson, X. Zhang, H. Zhang, et al. 2016. RNA polymerase II regulates topoisomerase 1 activity to favor efficient transcription. *Cell.* 165:357–371.

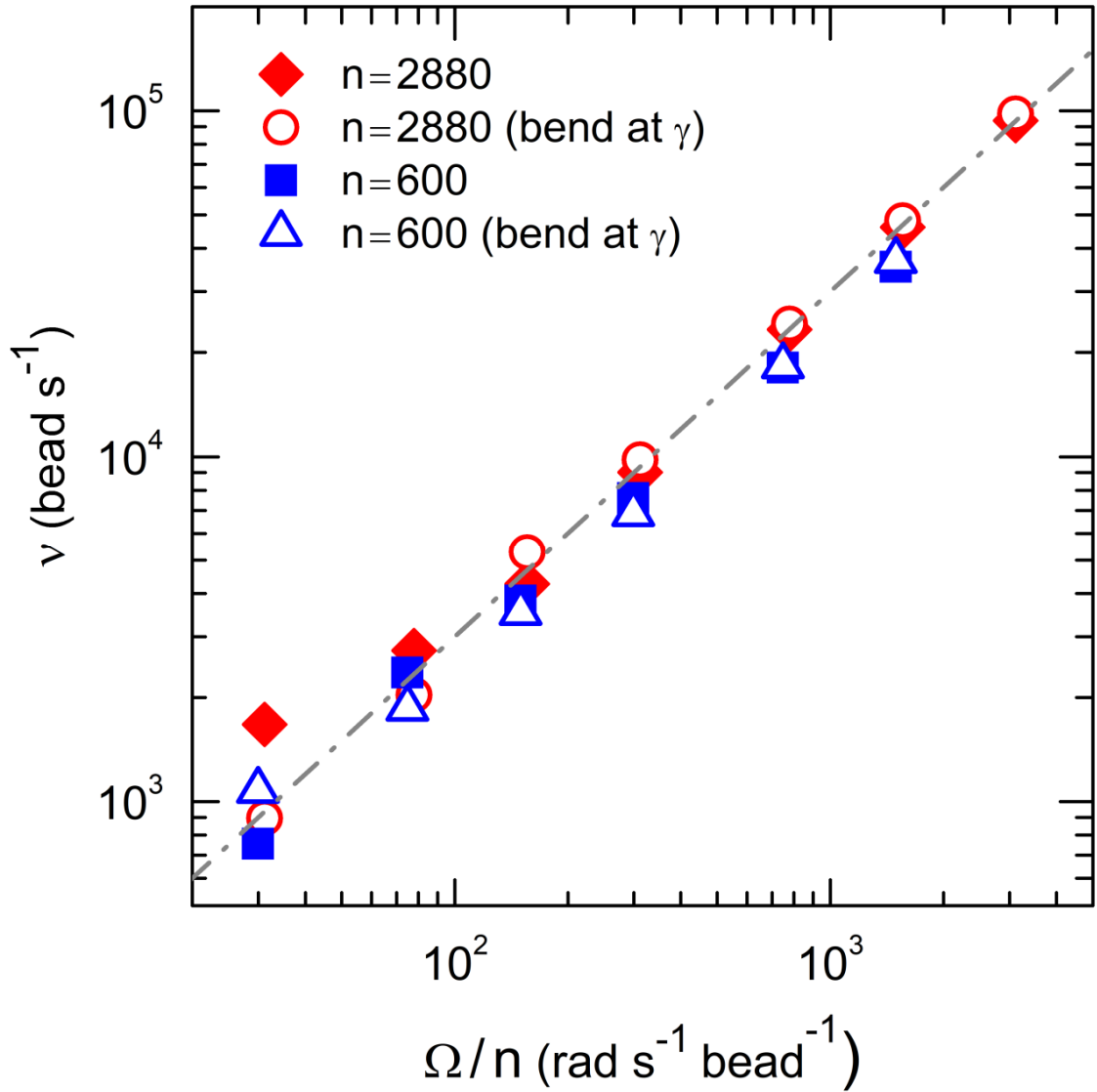

**Figure 1.** Log-log plot of the evolution of $\nu$ as a function of $\Omega/n$ in the absence of topological barrier. Blue (respectively, red) symbols correspond to simulations performed with $n=600$ (respectively, $n=2880$) beads. Open (respectively, filled) symbols correspond to simulations performed with (respectively, without) a static bend of 40° at bead $\gamma$. The gray dot-dashed line has equation $\nu = 30\,(\Omega/n)$ and is a guideline for the eyes.

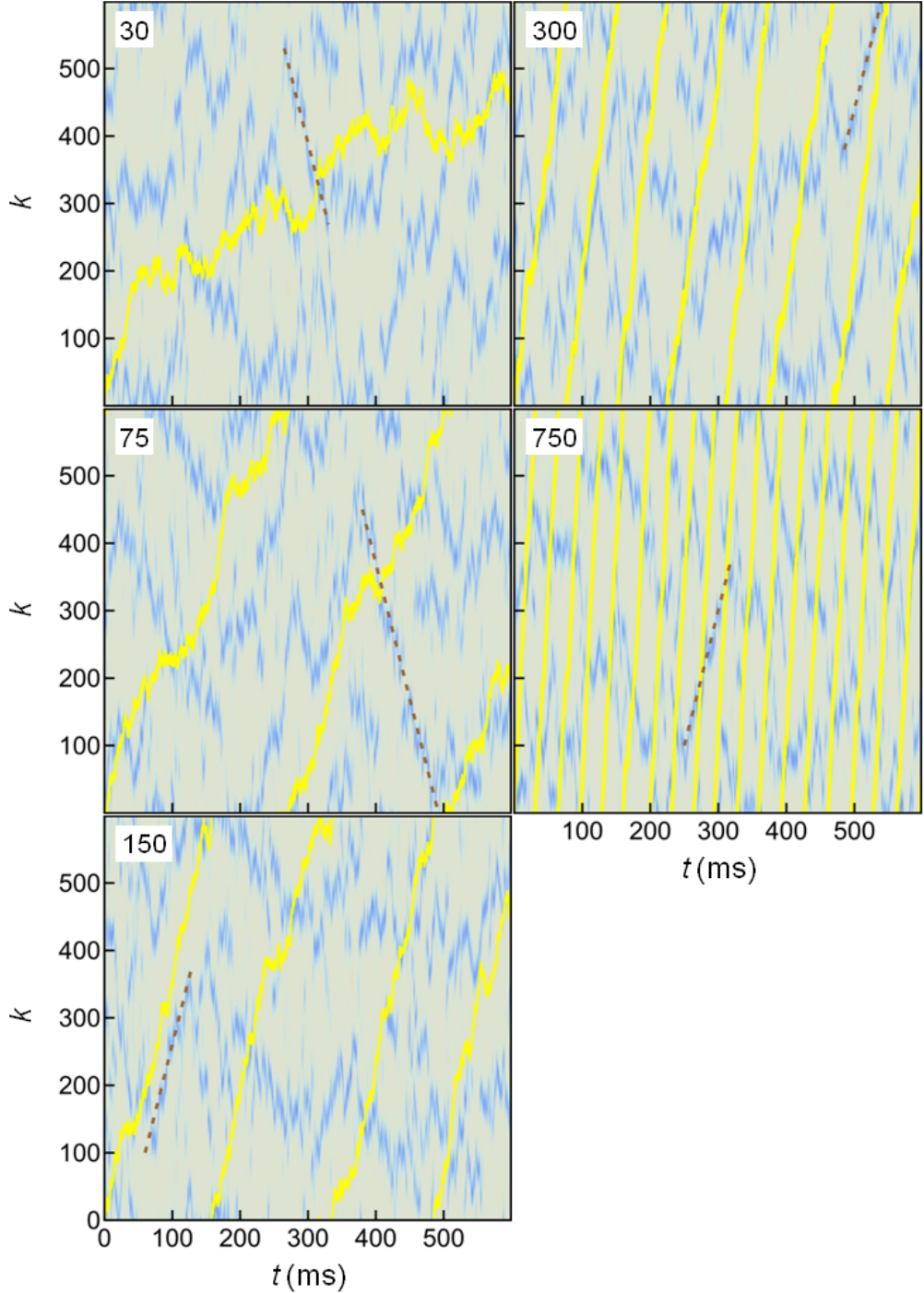

**Figure 2.** Time evolution of the local writhe $Wr(k)$ for circular DNA chains with $n=600$ beads and values of $\Omega/n$ ranging from 30 to 750 rad×s$^{-1}$×bead$^{-1}$, as indicated in the top left corner of each panel. Simulations were performed without static bend at bead $\gamma$. The value of $Wr(k)$ at DNA bead $k$ and time $t$ is represented according to a color code ranging from deep blue ($Wr(k) \leq -3$) to deep red ($Wr(k) \geq 3$). The yellow lines indicate the position of the motor at each time step. The brown dashed lines are guidelines for the eyes, which highlight portions of trajectories with identical slithering velocity of 4000 beads per second. Remember that the DNA chain is circular and that bead $k=1$ is bound to bead $k=600$.

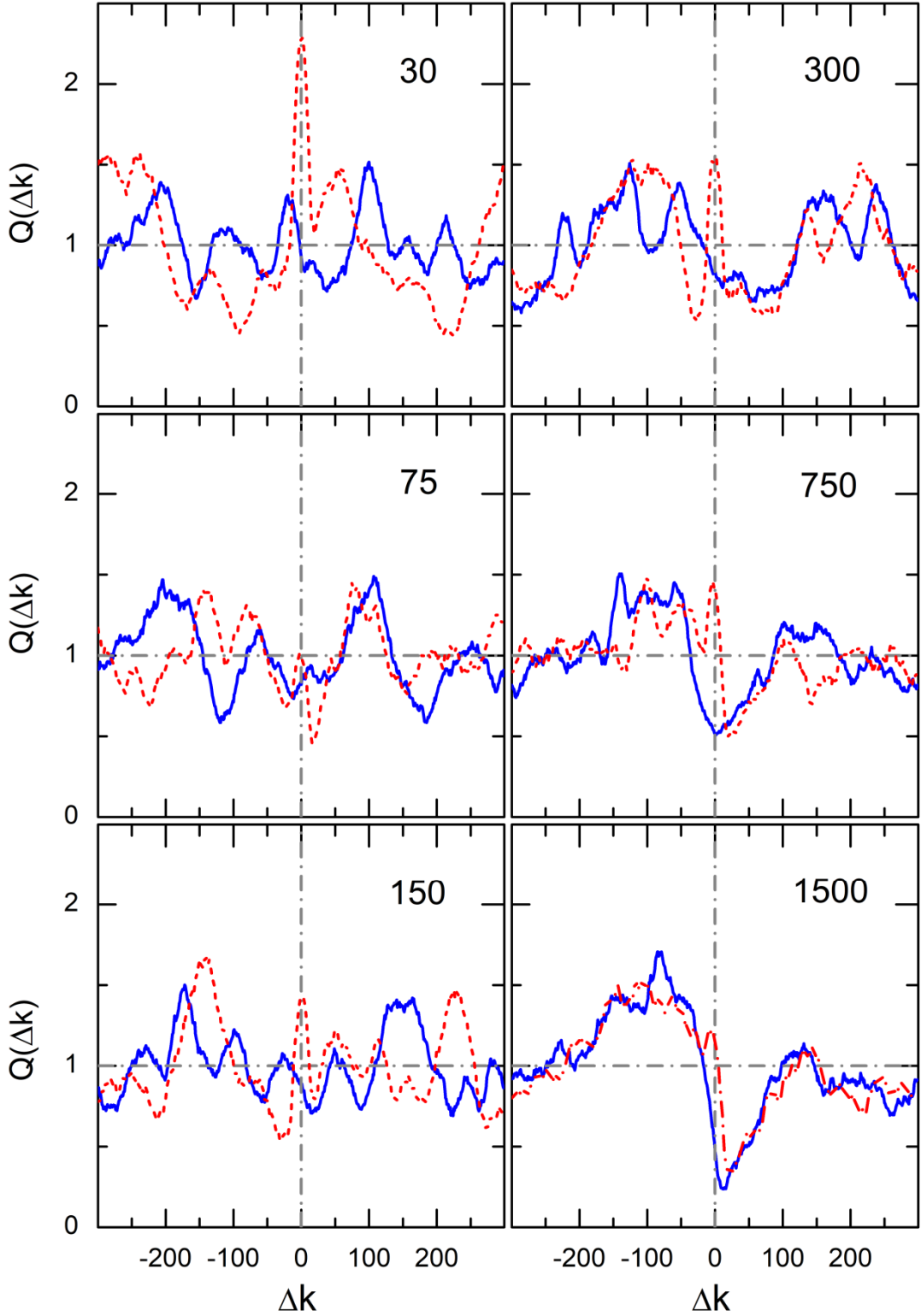

**Figure 3.** Plot of the probability $Q(\Delta k)$ that the local writhe at DNA bead $\gamma + \Delta k$ satisfies $Wr(\gamma + \Delta k) \leq -1.4$, for circular DNA chains with $n = 600$ beads and values of $\Omega/n$ ranging from 30 to 750 rad×s$^{-1}$×bead$^{-1}$, as indicated in the top right corner of each panel. Solid blue (respectively, dashed red) lines correspond to simulations performed without (respectively, with) a static bend of 40° at bead $\gamma$.

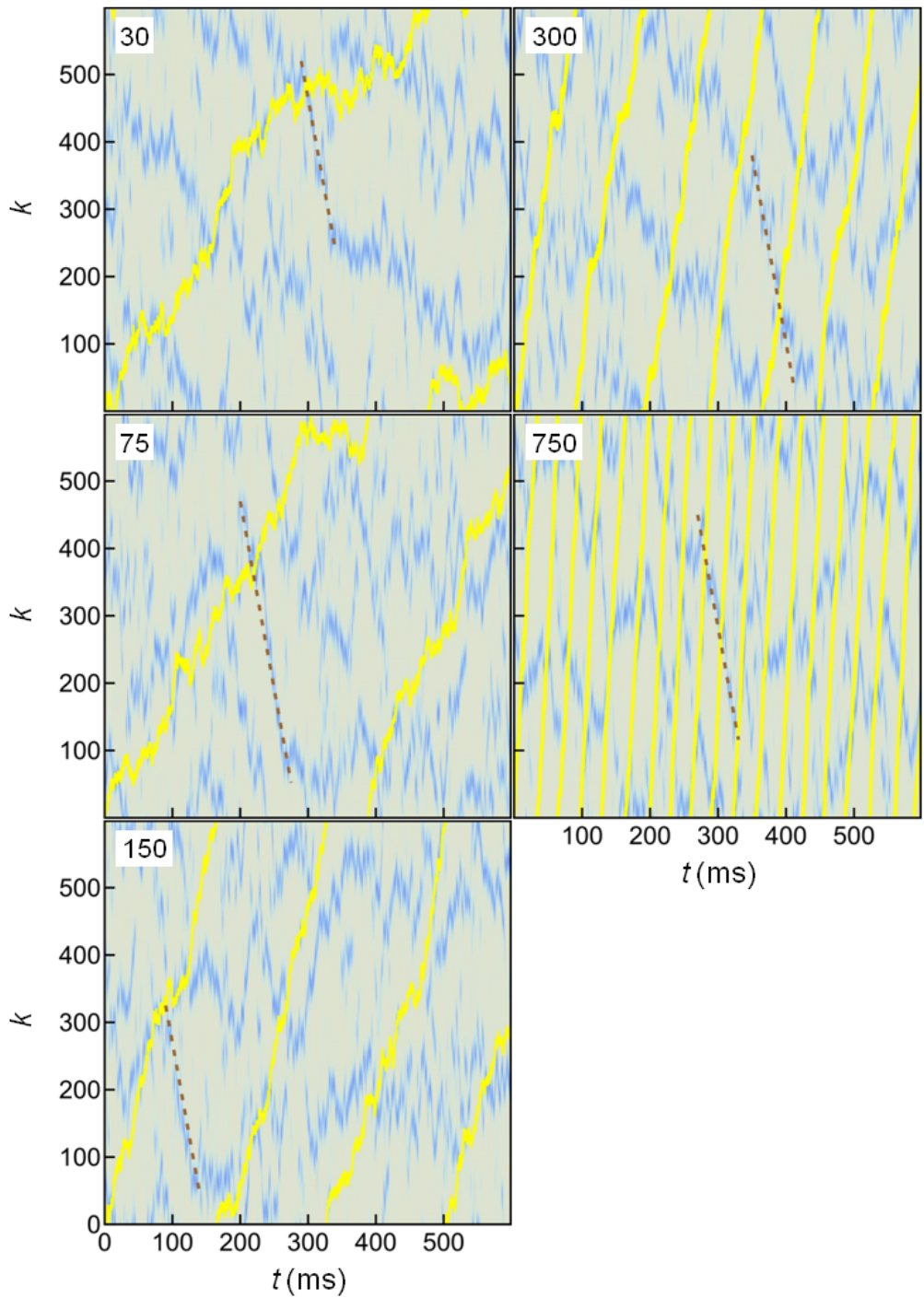

**Figure 4.** Same as Fig. 2, except that simulations were performed with a static bend of 40° at bead $\gamma$ and that brown dashed lines highlight portions of trajectories with slithering velocity of 5600 beads per second.

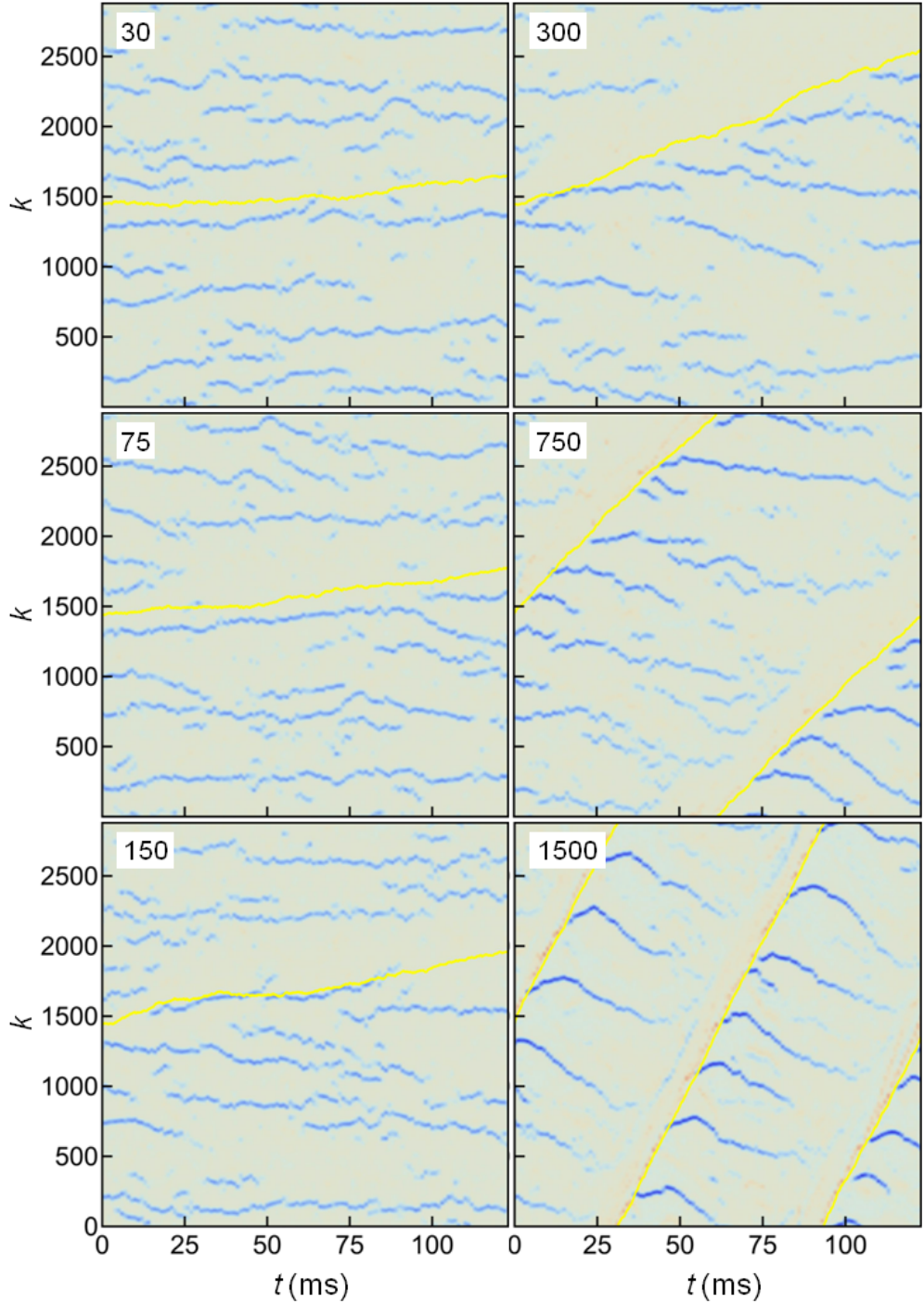

**Figure 5.** Time evolution of the local writhe $Wr(k)$ for circular DNA chains with $n = 2880$ beads and values of $\Omega/n$ ranging from 30 to 1500 rad×s$^{-1}$×bead$^{-1}$, as indicated in the top left corner of each panel. Simulations were performed without static bend at bead $\gamma$. The value of $Wr(k)$ at DNA bead $k$ and time $t$ is represented according to a color code ranging from deep blue ($Wr(k) \leq -3$) to deep red ($Wr(k) \geq 3$). The yellow lines indicate the position of the motor at each time step.

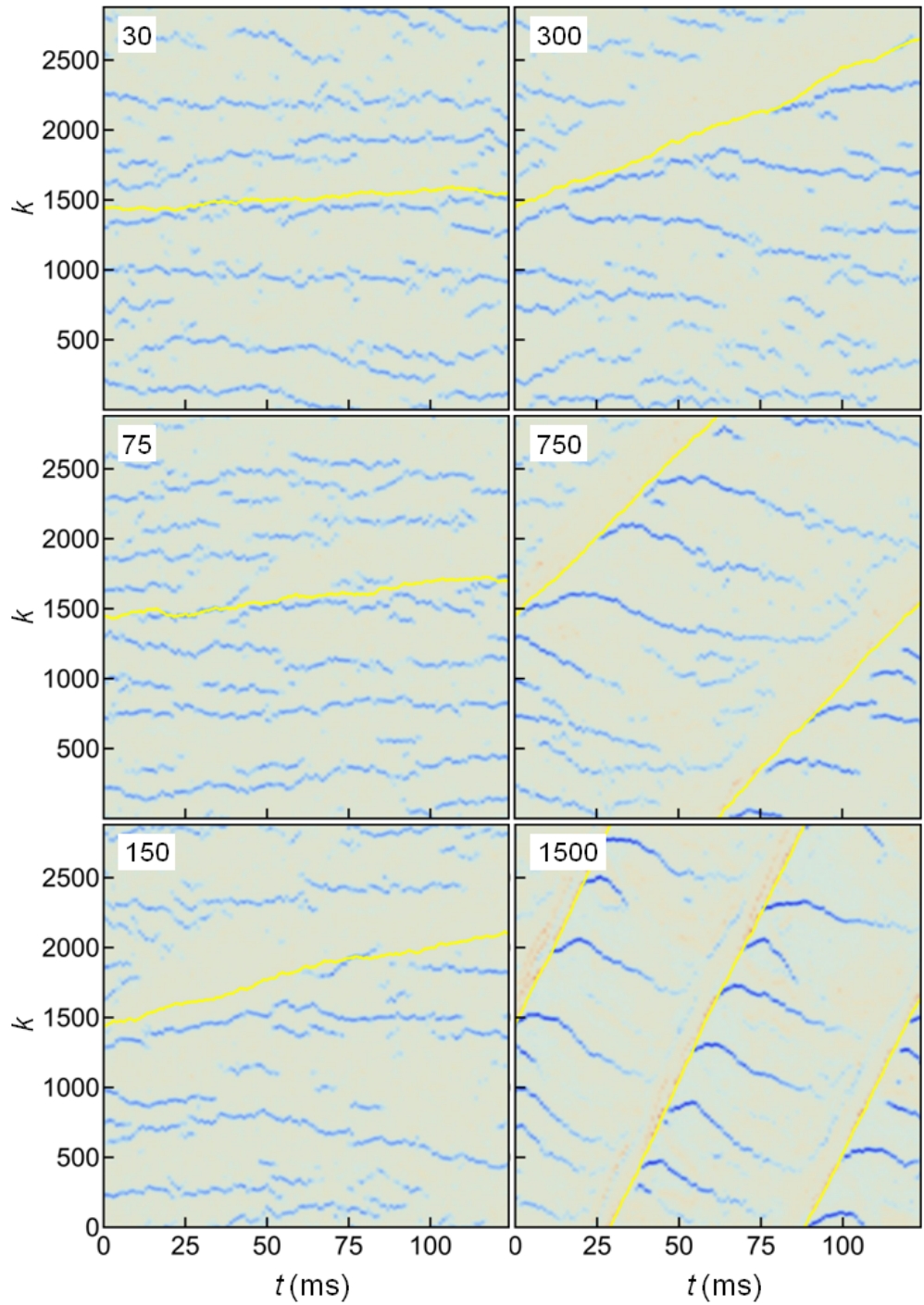

**Figure 6.** Same as Fig. 5, except that simulations were performed with a static bend of 40° at bead $\gamma$.

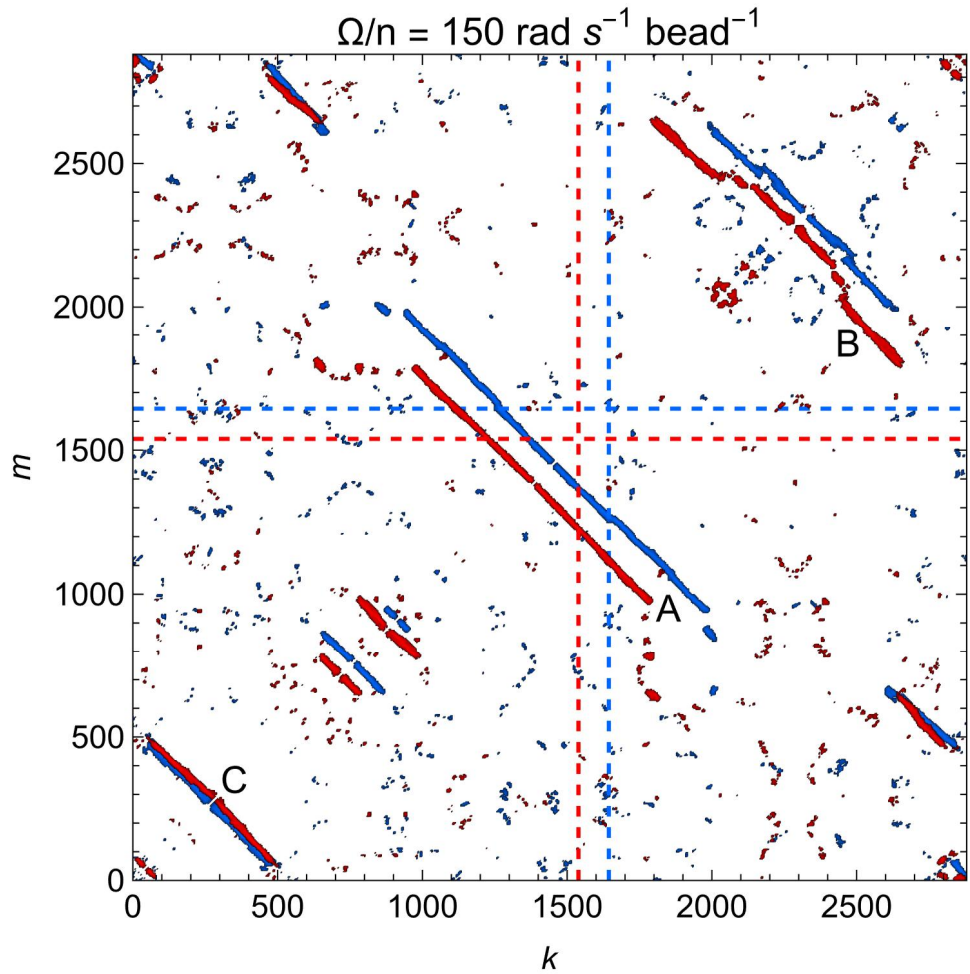

**Figure 7.** Superposed contact maps extracted at $t = 15$ ms (red) and $t = 35$ ms (blue) from the simulation with $\Omega/n = 150$ rad×s$^{-1}$×bead$^{-1}$ shown in Fig. 6. Plectonemes appear as segments parallel to the secondary diagonal. The red and blue dot-dashed lines show the respective locations of the motor.

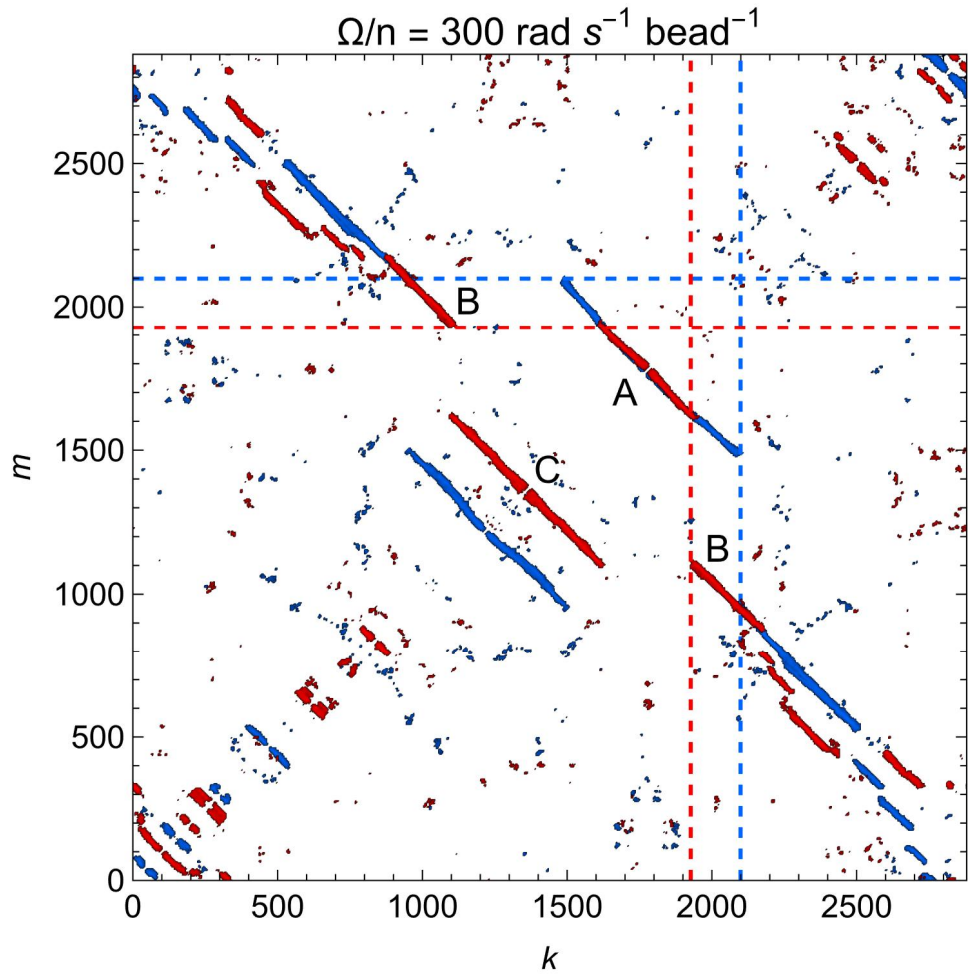

**Figure 8.** Superposed contact maps extracted at $t = 51$ ms (red) and $t = 71$ ms (blue) from the simulation with $\Omega/n = 300$ rad×s$^{-1}$×bead$^{-1}$ shown in Fig. 6. Plectonemes appear as segments parallel to the secondary diagonal. The red and blue dot-dashed lines show the respective locations of the motor.

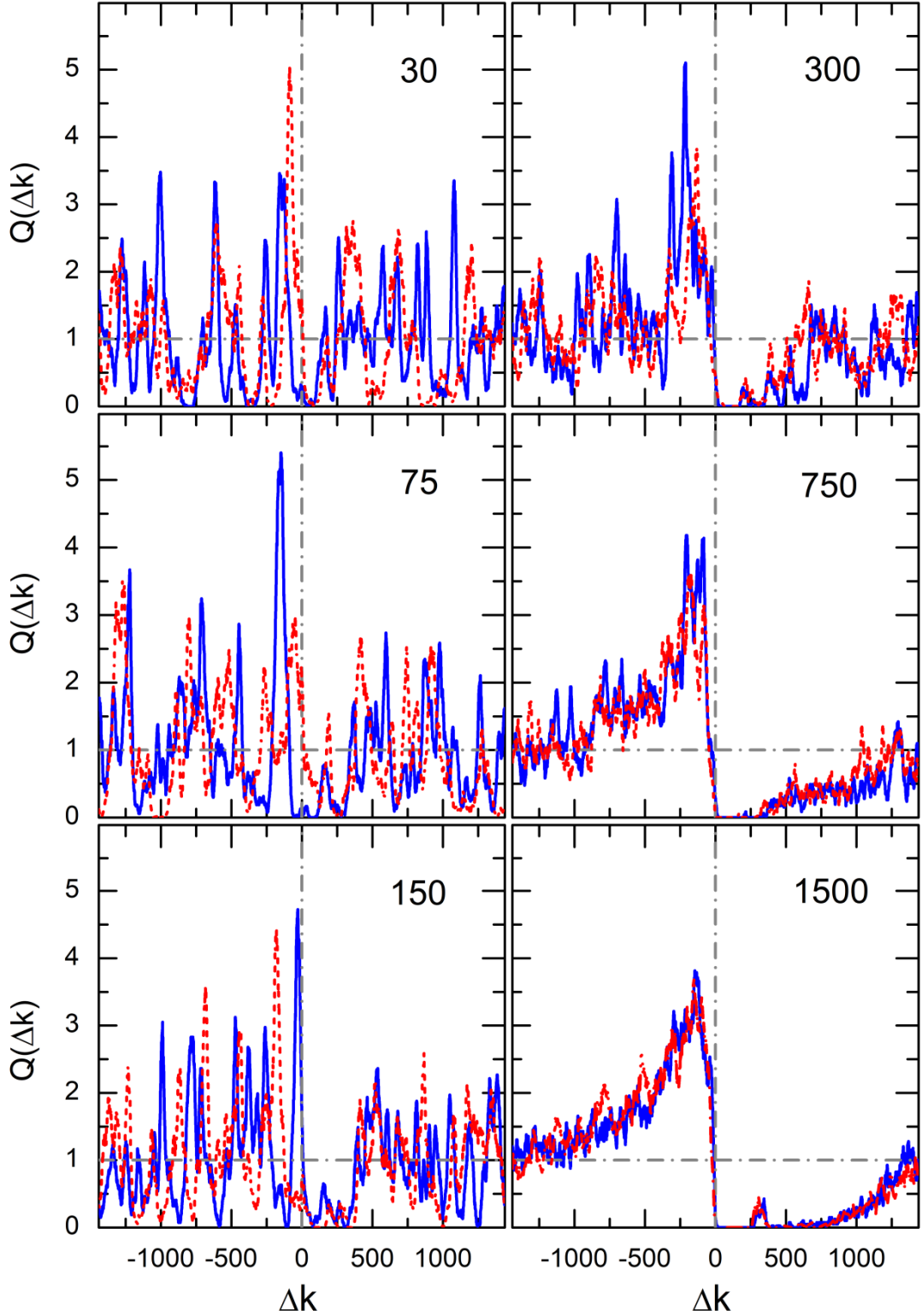

**Figure 9.** Plot of the probability $Q(\Delta k)$ that the local writhe at DNA bead $\gamma + \Delta k$ satisfies $Wr(\gamma + \Delta k) \leq -1.4$, for circular DNA chains with $n = 2880$ beads and values of $\Omega/n$ ranging from 30 to 1500 rad×s$^{-1}$×bead$^{-1}$, as indicated in the top right corner of each panel. Solid blue (respectively, dashed red) lines correspond to simulations performed without (respectively, with) a static bend of 40° at bead $\gamma$.

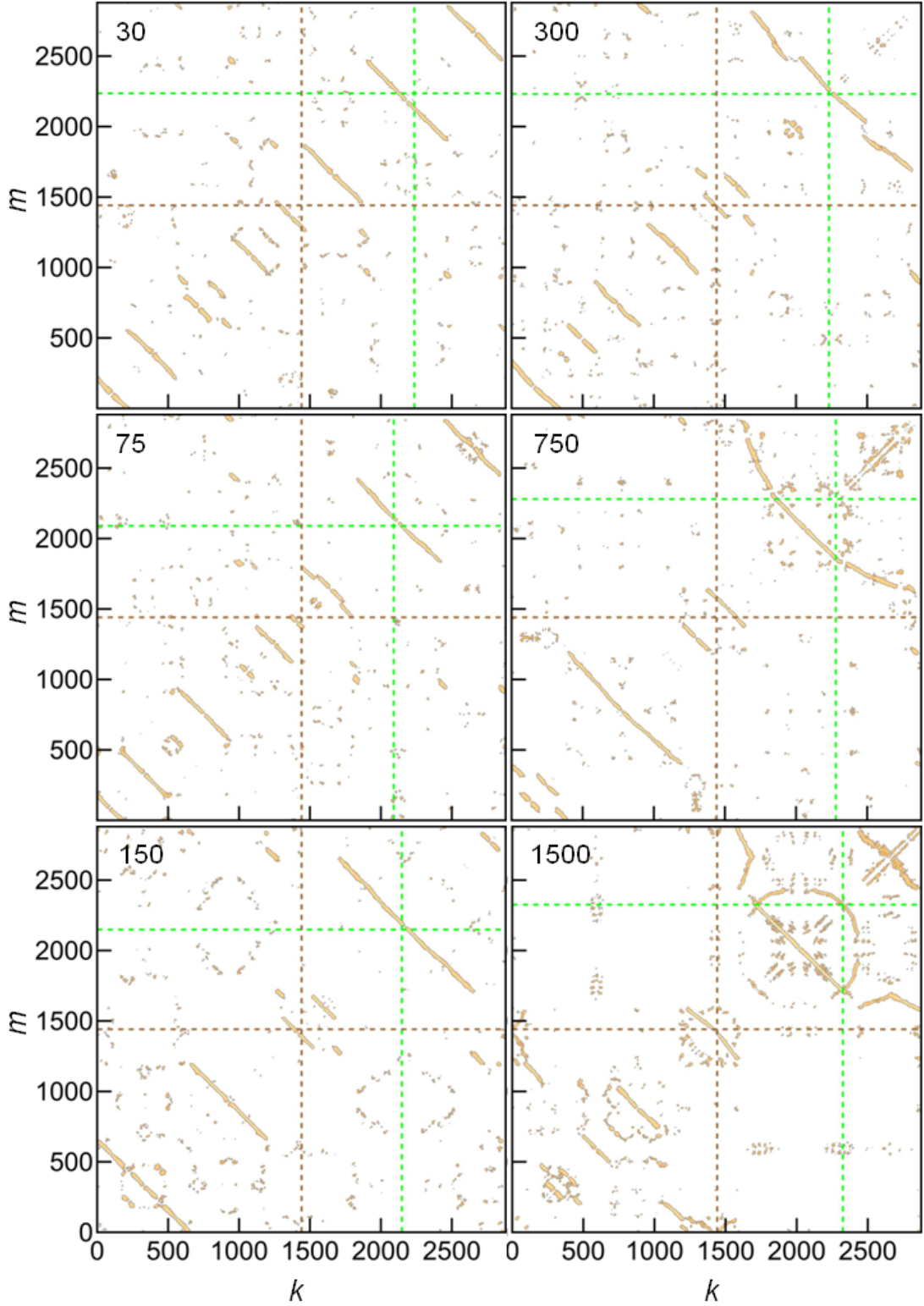

**Figure 10.** Contact maps extracted at $t = 50$ ms from simulations performed with a topological barrier that bridges beads $\alpha = 1$ and $\beta = 1441$, a static bend of 40° at bead $\gamma$, and values of $\Omega / n$ ranging from 30 to 1500 rad×s$^{-1}$×bead$^{-1}$, as indicated in the top left corner of each panel. Plectonemes appear as segments parallel to the secondary diagonal. Dashed brown lines indicate the position of bead $\beta$, dashed green lines the position of the motor.

**TOC Graphic**

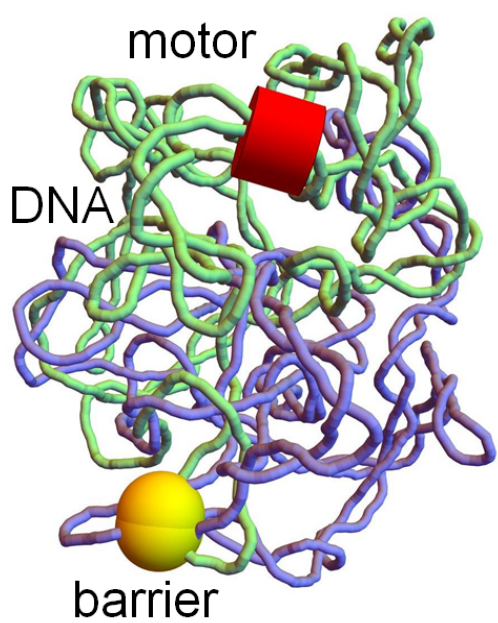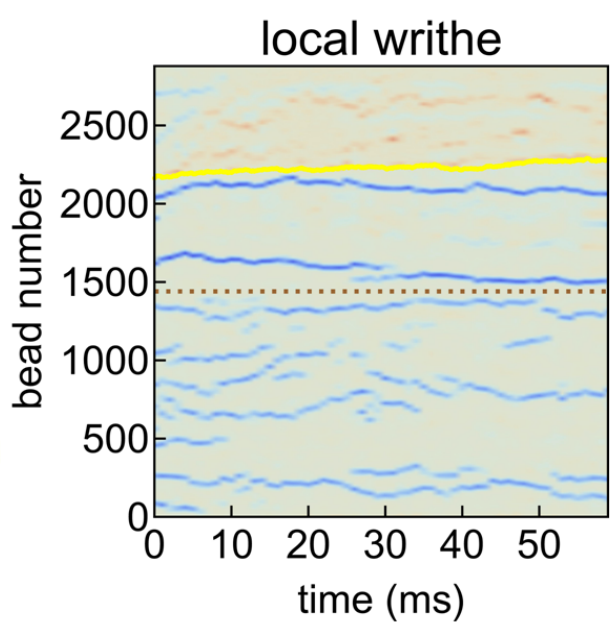

**Twin Supercoil Domain Couples the Dynamics of Molecular Motors and Plectonemes During Bacterial DNA Transcription and Replication**

Marc JOYEUX [*]

*Laboratoire Interdisciplinaire de Physique,*
*CNRS and Université Grenoble Alpes,*
*38400 St Martin d'Hères,*
*France*

**SUPPORTING INFORMATION**

Figures S1 to S5, showing representative snapshots extracted from the simulations, additional plots of the local writhe $Wr(k)$, and additional contact maps.

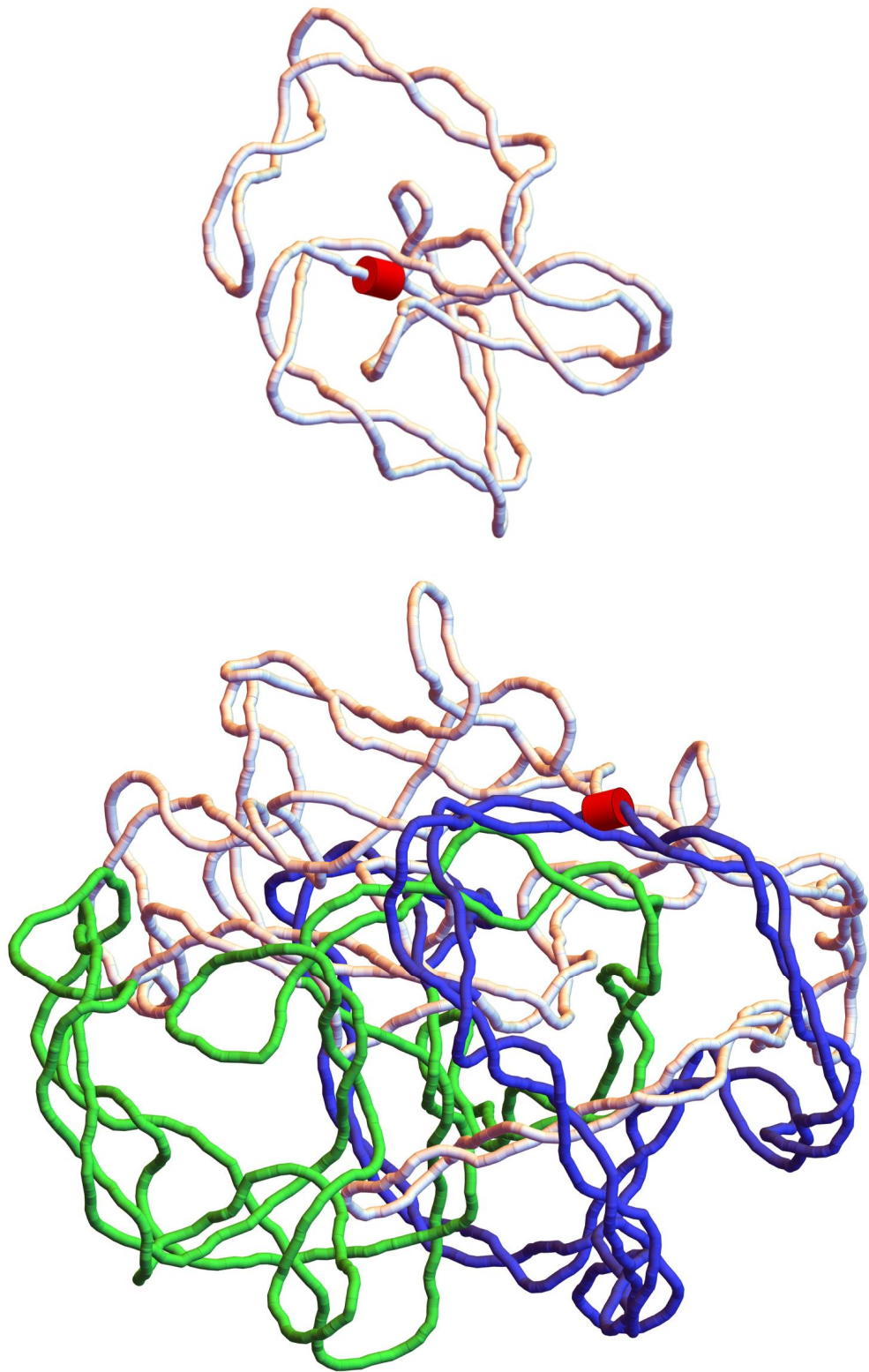

**Figure S1.** Snapshots extracted from simulations with $\Omega/n = 150$ rad×s$^{-1}$×bead$^{-1}$, a static bend of 40° at bead $\gamma$, and $n = 600$ beads (top panel) or $n = 2880$ beads (bottom panel). The DNA chain is represented as a tube and the motor as a red cylinder.

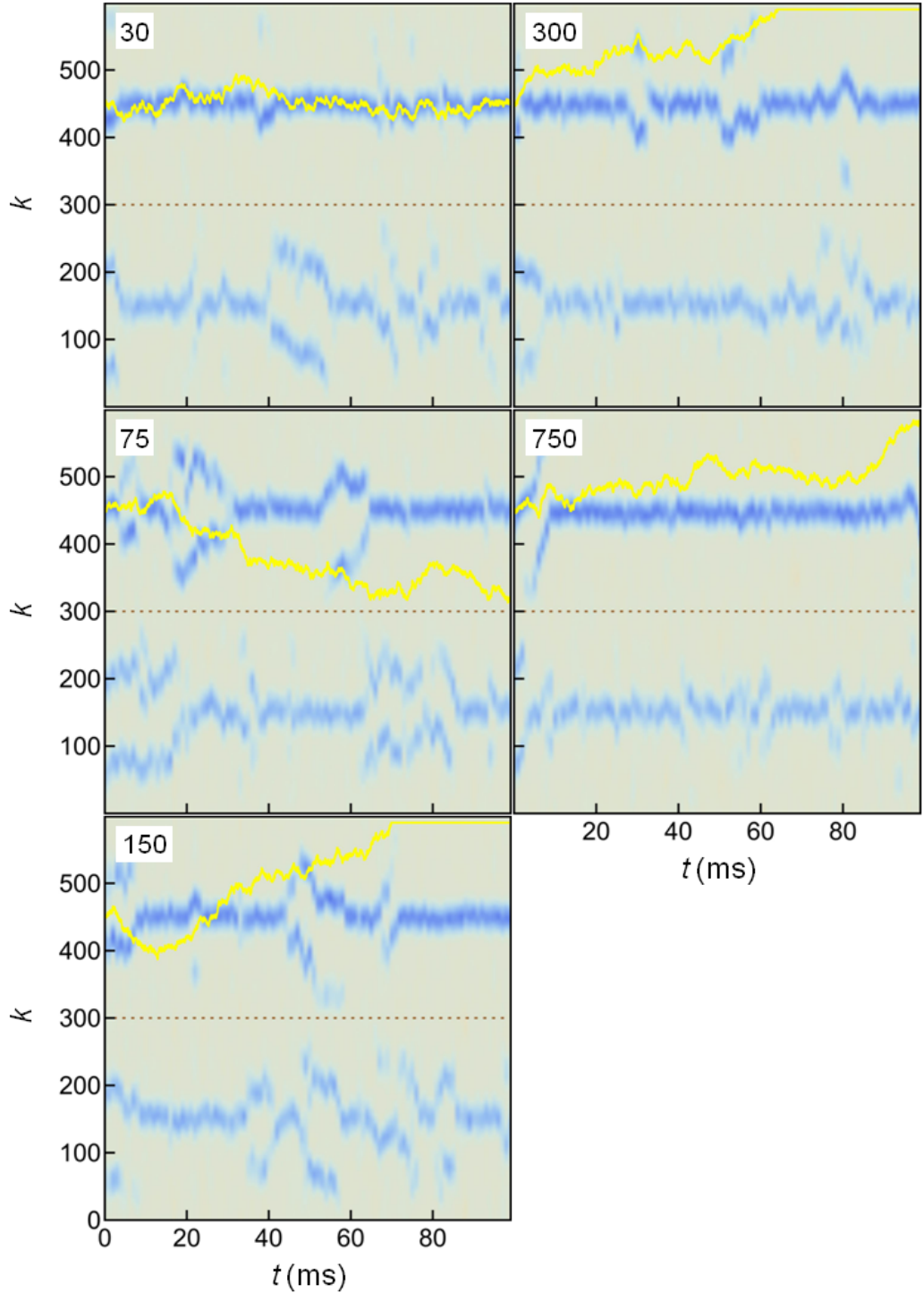

**Figure S2.** Time evolution of the local writhe $Wr(k)$ for circular DNA chains with $n = 600$ beads obtained from simulations performed with a topological barrier that bridges beads $\alpha = 1$ and $\beta = 301$, a static bend of 40° at bead $\gamma$, and values of $\Omega/n$ ranging from 30 to 750 rad×s$^{-1}$×bead$^{-1}$, as indicated in the top left corner of each panel. The value of $Wr(k)$ at DNA bead $k$ and time $t$ is represented according to a color code ranging from deep blue ($Wr(k) \leq -3$) to deep red ($Wr(k) \geq 3$). The yellow lines indicate the position of the motor at each time step, the brown dashed lines the position of bead $\beta$.

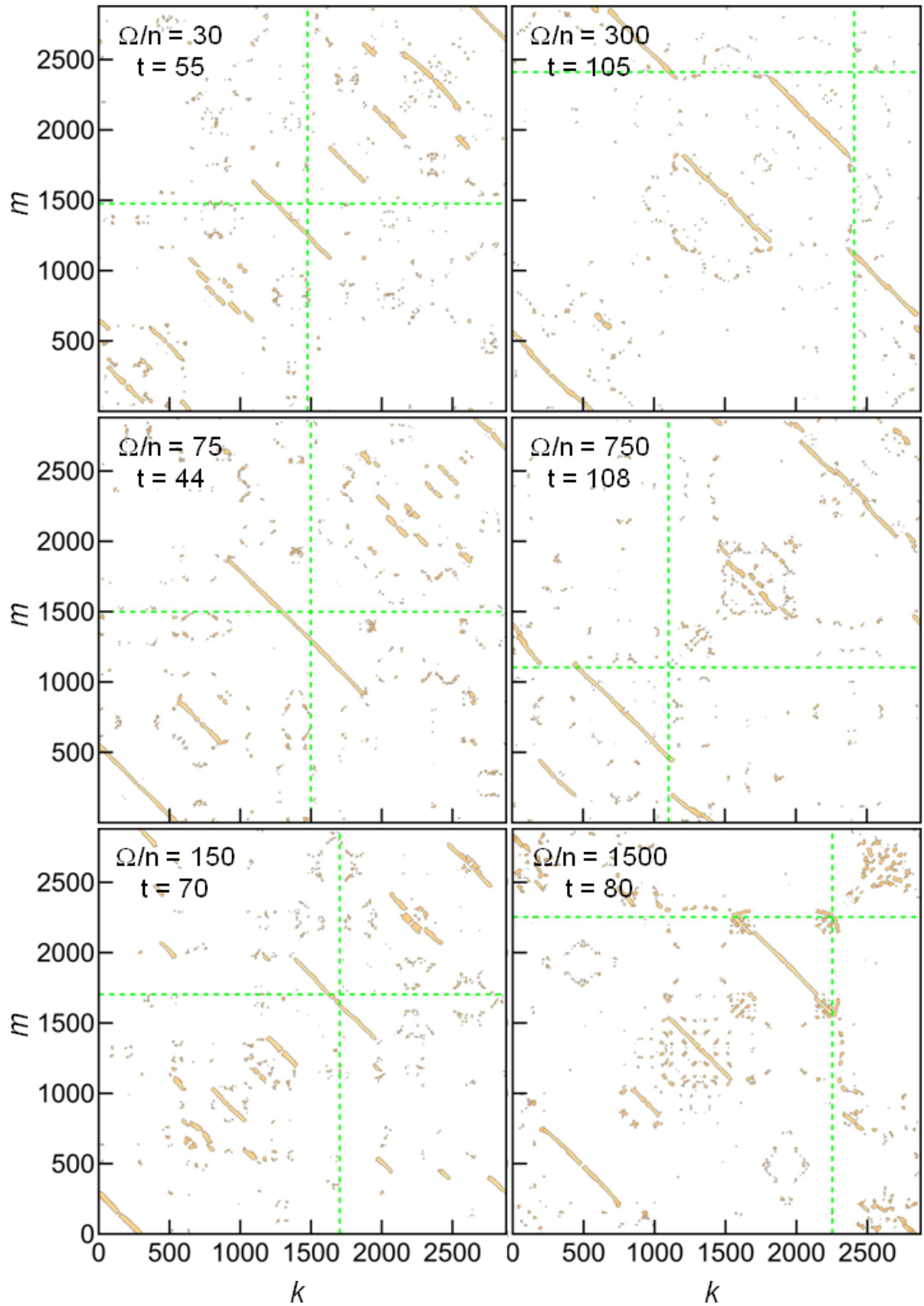

**Figure S3.** Representative contact maps extracted from the simulations shown in Fig. 5. The values of $\Omega/n$ (expressed in rad×s$^{-1}$×bead$^{-1}$) and time $t$ (expressed in ms), are indicated in the top left corner of each panel. Plectonemes appear as segments parallel to the secondary diagonal.

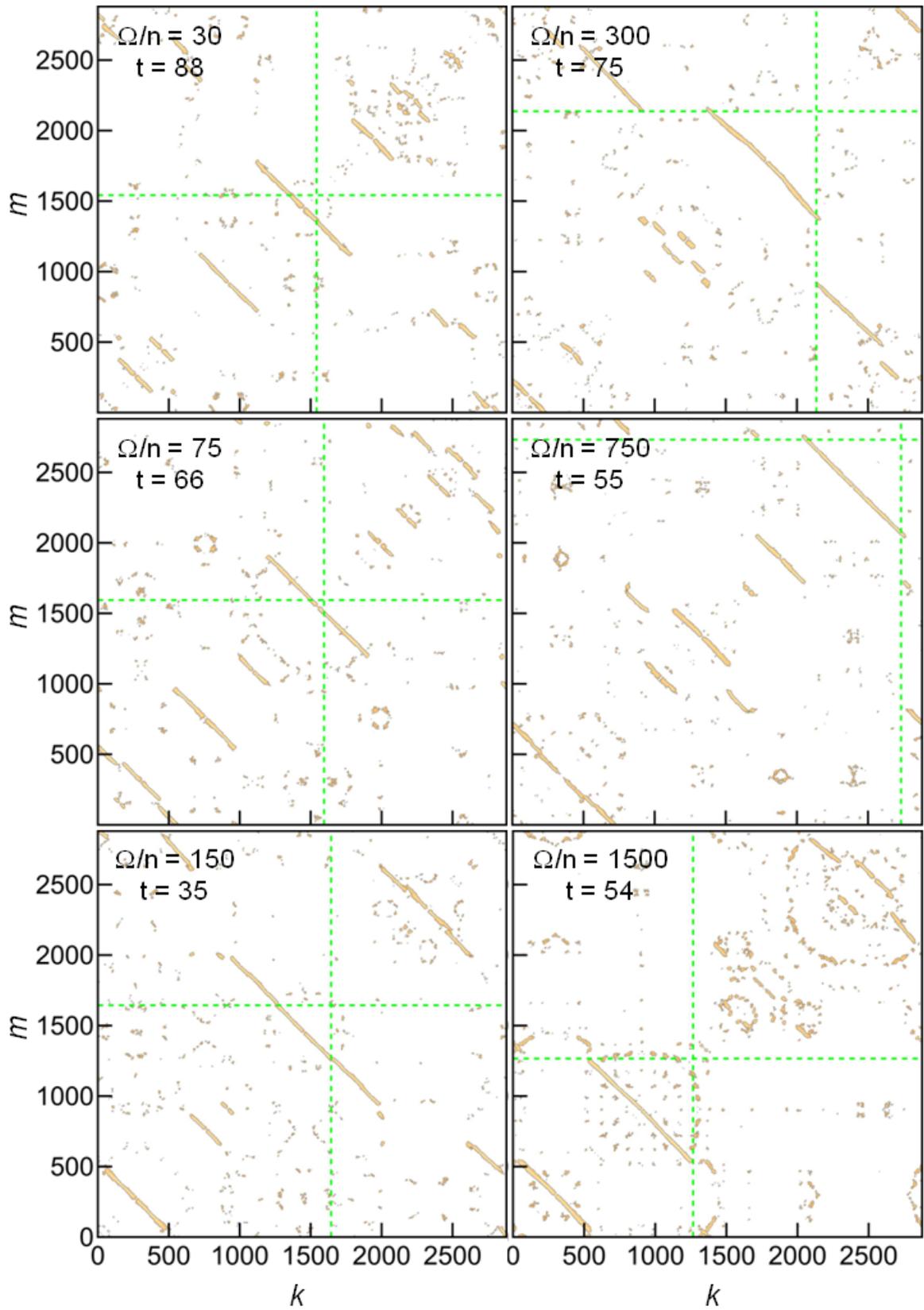

**Figure S4.** Representative contact maps extracted from the simulations shown in Fig. 6. The values of $\Omega/n$ (expressed in rad×s$^{-1}$×bead$^{-1}$) and time $t$ (expressed in ms), are indicated in the top left corner of each panel. Plectonemes appear as segments parallel to the secondary diagonal.

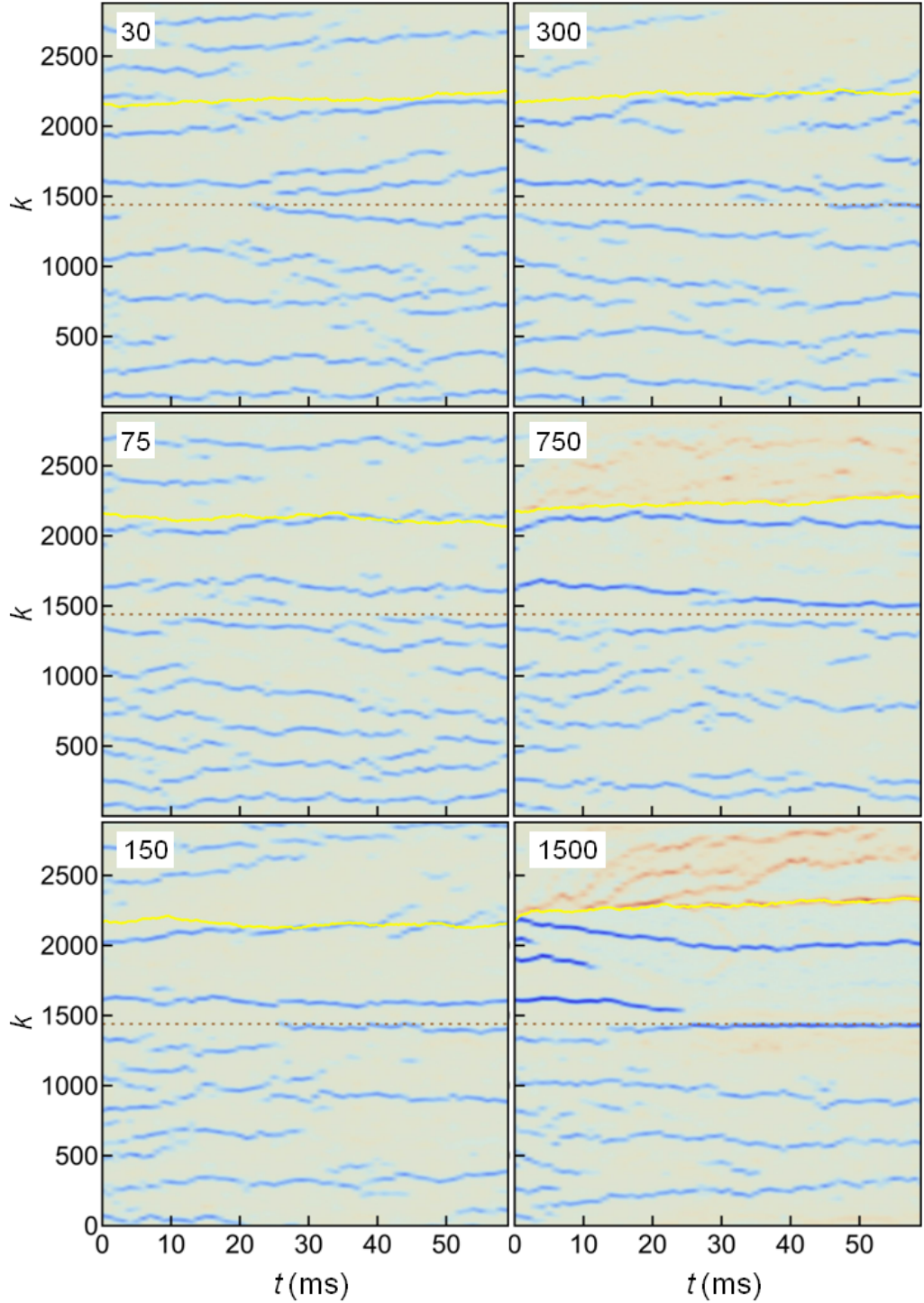

**Figure S5.** Time evolution of the local writhe $Wr(k)$ for circular DNA chains with $n = 2880$ beads obtained from simulations performed with a topological barrier that bridges beads $\alpha=1$ and $\beta=1441$, a static bend of 40° at bead $\gamma$, and values of $\Omega/n$ ranging from 30 to 1500 rad×s$^{-1}$×bead$^{-1}$, as indicated in the top left corner of each panel. The value of $Wr(k)$ at DNA bead $k$ and time $t$ is represented according to a color code ranging from deep blue ($Wr(k) \leq -3$) to deep red ($Wr(k) \geq 3$). The yellow lines indicate the position of the motor at each time step, the brown dashed lines the position of bead $\beta$.